\documentclass[useAMS]{mn2e}
\usepackage{times}
\input amssym.def
\input amssym.tex

\makeatletter
\let\jnl@style=\rmfamily 
\def\ref@jnl#1{{\jnl@style#1}}% 
\newcommand\aj{\ref@jnl{AJ}}% 
\newcommand\apj{\ref@jnl{ApJ}}% 
\newcommand\apjl{\ref@jnl{ApJ}}% 
\newcommand\apjs{\ref@jnl{ApJS}}% 
\newcommand\aap{\ref@jnl{A\&A}}% 
\newcommand\aaps{\ref@jnl{A\&AS}}% 
\newcommand\mnras{\ref@jnl{MNRAS}}% 
\newcommand\pasp{\ref@jnl{PASP}}% 
\makeatother

\newcommand\nodata{ ~$\cdots$~ }% 

\usepackage{graphicx}

\makeatletter
\newcommand\ion[2]{#1$\;${\small\rmfamily\@Roman{#2}}\relax}%
\newcommand{\cion}[2]{\ion{#1}{#2}}
\newcommand{\ionl}[3]{\mbox{#1$\;${\sc\@roman{#2}}$\,\lambda\,$#3\relax}}
\makeatother

\newcommand{\fcion}[2]{\mbox{$f$(\ion{#1}{#2})}}
\newcommand{\Nion}[2]{\mbox{$N$(\ion{#1}{#2})}}

\def\hs0624{\mbox{HS~0624+6907}\relax}
\def\lya{\mbox{Ly$\alpha$}\relax}
\def\lyb{\mbox{Ly$\beta $}\relax}
\def\lyg{\mbox{Ly$\gamma$}\relax}

\def\fig#1{Figure \ref{#1}}

\newcommand{\abell}[1]{A#1\relax} 
\newcommand{\labell}[1]{A#1\relax} 

\newcommand{\zapp}[1]{\mbox{$z\approx#1$}\relax}
\newcommand{\zabsapp}[1]{\mbox{$z_{\rm abs}\approx#1$}\relax}
\newcommand{\zbtw}[2]{\mbox{$#1\leq z \leq#2$}\relax}
\newcommand{\zabs}[1]{\mbox{$z_{\rm abs}=#1$}\relax}
\def\zva{0.064}
\def\zvb{0.077}
\def\zvc{0.110}

\def\hmMpc{\mbox{$h_{70}^{-1}\,{\rm Mpc}$}}
\def\hmkpc{\mbox{$h_{70}^{-1}\,{\rm kpc}$}}
\def\kms{\mbox{km~s$^{-1}$}\relax}
\def\cmmd{\mbox{cm$^{-2}$}\relax}
\def\FUSE{{\it FUSE}}

\def\colhead#1{#1}
\def\startdata{\\\hline}
\def\enddata{\\\hline}
\def\tableminisize{150mm}
\newcommand\phn{\phantom{0}}

\newcommand\tablenotemark[1]{\mbox{$^#1$}}
\newcommand\tablenotetext[2]{\footnote{#2}}

\begin{document}

%\title{Physical Properties of the Gaseous Cosmic Web: High
%Metallicity, Photoionised Gas toward HS~0624+6907\altaffilmark{1}}
\title[High Metallicity, Photoionised Gas in Intergalactic Large-Scale Filaments]{High Metallicity, Photoionised Gas in Intergalactic Large-Scale Filaments\thanks{Based on observations with (1) the NASA/ESA {\it Hubble Space Telescope}, obtained at the Space Telescope Science
Institute, which is operated by the Association of Universities for
Research in Astronomy, Inc., under NASA contract NAS 5-26555, (2) the
NASA-CNES/ESA {\it Far Ultraviolet Spectroscopic Explorer} mission,
operated by Johns Hopkins University, supported by NASA contract NAS
5-32985, and (3) the Apache Point Observatory 3.5m telescope, which is
owned and operated by the Astrophysical Research Consortium.}}

\author[B. Aracil et al.]{Bastien Aracil,$^1$ Todd M. Tripp,$^{1}$\thanks{Visiting Astronomer, Kitt Peak National Observatory, National Optical Astronomy Observatory, which is operated by the
Association of Universities for Research in Astronomy, Inc. (AURA)
under cooperative agreement with the National Science Foundation.} David V. Bowen,$^2$ Jason X. Prochaska,$^3$
\newauthor
Hsiao-Wen Chen$^4$ and Brenda L. Frye$^2$\\
$^1$Department of Astronomy, University of Massachusetts, 710 N. Pleasant St., Amherst, MA 01003-9305\\
$^2$Princeton University Observatory, Peyton Hall, Princeton, NJ 08544\\
$^3$University of California Observatories, Natural Sciences II Annex, University of California, Santa Cruz, CA 95064\\
$^4$Department of Astronomy \& Astrophysics, University of Chicago, Chicago, IL 60637 }

\maketitle

\begin{abstract}

We present high-resolution ultraviolet spectra of absorption-line
systems toward the low$-z$ QSO \hs0624 ($z_{\rm QSO} =
0.3700$). Coupled with ground-based imaging and spectroscopic galaxy
redshifts, we find evidence that many of these absorbers do not arise
in galaxy halos but rather are truly integalactic gas clouds
distributed within large-scale structures, and moreover, the gas is
cool ($T < 10^{5}$ K) and has relatively high metallicity ($Z = 0.9
Z_{\odot}$).  {\it HST} Space Telescope Imaging Spectrograph (STIS)
data reveal a dramatic cluster of 13 \ion{H}{1} \lya\ lines within a
1000 \kms\ interval at $z_{\rm abs} = 0.0635$. We find 10 galaxies at
this redshift with impact parameters ranging from $\rho = 135
h_{70}^{-1}$ kpc to 1.37 $h_{70}^{-1}$ Mpc. The velocities and
velocity spread of the \lya\ lines in this complex are unlikely to
arise in the individual halos of the nearby galaxies; instead, we
attribute the absorption to intragroup medium gas, possibly from a
large-scale filament viewed along its long axis.  Contrary to
theoretical expectations, this gas is not the shock-heated warm-hot
intergalactic medium (WHIM); the width of the \lya\ lines all indicate
a gas temperature $T \ll 10^{5}$ K, and metal lines detected in the
\lya\ complex also favor photoionised, cool gas.  No \ion{O}{6}
absorption lines are evident, which is consistent with photoionisation
models.  Remarkably, the metallicity is near-solar, [M/H] $= -0.05 \pm
0.4$ ($2\sigma$ uncertainty), yet the nearest galaxy which might
pollute the IGM is at least 135~$h_{70}^{-1}$ kpc away.  Tidal
stripping from nearby galaxies appears to be the most likely origin of
this highly enriched, cool gas.  More than six Abell galaxy clusters
are found within $4^{\circ}$ of the sight line suggesting that the QSO
line of sight passes near a node in the cosmic web.  At $z \approx$
0.077, we find absorption systems as well as galaxies at the redshift
of the nearby clusters Abell 564 and Abell 559. We conclude that the
sight line pierces a filament of gas and galaxies feeding into these
clusters.  The absorber at $z_{\rm abs}$ = 0.07573 associated with
Abell 564/559 also has a high metallicity with [C/H] $> -0.6$, but
again the closest galaxy is relatively far from the sight line ($\rho
= 293\,h_{70}^{-1}$ kpc).  The Doppler parameters and \ion{H}{1}
column densities of the Ly$\alpha$ lines observed along the entire
sight line are consistent with those measured toward other low$-z$
QSOs, including a number of broad ($b>40$~\kms) \lya\ lines.

\end{abstract}


\section{Introduction}

In cold  dark matter cosmology,  the initially smooth  distribution of
matter in the universe is  expected to collapse into a complex network
of filaments and voids, structures which have been termed the ``cosmic
web''. The filamentary distribution of galaxies in the nearby universe
has been  revealed in detail  by recent large galaxy  redshift surveys
such as  the 2dFGRS (Colless et  al.  2001, Baugh et  al.  2004), the
Sloan Digital  Sky Survey (SDSS, Stoughton et  al.  2002, Doroshkevich
et al.  2004) and the  2$\mu m$ All  Sky Survey (2MASS, Maller  et al.
2002).  Numerical  simulations  successfully  reproduce  this  network
(Jenkins et al. 1998; Colberg  et al. 2004) and indicate that galaxies
are only the tip of the iceberg  in this cosmic web (Katz et al. 1996;
Miralda-Escud\'{e}  et al.  1996).   Hydrodynamic simulations  suggest
that at  the present epoch, in  addition to dark  matter and galaxies,
the filaments are also composed of a mixture of cool, photoionised gas
(the  low$-z$  remnants of  the  \lya\  forest)  and a  shock  heated,
low-density  gaseous  phase   at  temperatures  between  $10^5$~K  and
$10^7$~K  that contains most  of the  baryonic mass,  the ``warm-hot''
intergalactic  medium  (WHIM,  Cen   \&  Ostriker  1999;  Dav\'{e}  et
al. 1999).

Observational constraints  on the physical  conditions, distribution,a
nd  metal  enrichment  of  gas  in the  low-redshift  cosmic  web  are
currently quite  limited. The  existence of the  WHIM appears to  be a
robust prediction  of cosmological  simulations (Dav\'e et  al. 2001).
Thus,  observational efforts  are increasingly  being invested  in the
search  for WHIM  gas  and, more  generally,  the gaseous  filamentary
structures predicted by the models. Large-scale gaseous filaments have
been detected in X-ray emission (Wang et al. 1997; Scharf et al. 2000;
Tittley \& Henriksen 2001; Rines  et al 2001). However, X-ray emission
studies  with current  facilities  predominantly reveal  gas which  is
hotter  and denser  than  the WHIM;  this  X-ray emitting  gas is  not
expected to contain a substantial portion of the present-epoch baryons
(Dav\'{e} et al.  2001).  The  most promising method for observing the
WHIM in  the near term is  to search for  UV (\ion{O}{6}, \ion{Ne}{8})
and X-ray  (\ion{O}{7}, \ion{O}{8}, \ion{Ne}{9})  absorption lines due
to  WHIM gas  in the  spectra of  background QSOs/AGNs  (Tripp  et al.
2000, 2001; Savage et al.   2002,2005; Nicastro et al.  2002; Bergeron
et al.  2002; Richter et al.  2004; Sembach et al.  2004; Prochaska et
al. 2004; Danforth  \& Shull 2005).  While absorption  lines provide a
sensitive and  powerful probe of  the WHIM, the pencil-beam  nature of
the measurement along a sight  line provides little information on the
context  of  the absorption,  e.g.,  whether  the  lines arise  in  an
individual galaxy disk/halo, a  galaxy group, or lower-density regions
of a large-scale filament or void.

Thus, to understand the nature of highly ionised absorbers at low
redshifts, several groups are pursuing deep galaxy redshift surveys
and observations of QSOs behind well-defined galaxy groups or
clusters.  For example, to study gas located in large-scale filaments,
Bregman et al. (2004) have searched for absorption lines indicative of
the WHIM in regions between galaxy clusters/superclusters and have
identified some candidates.  In this paper, we carry out a similar
search as part of a broader program that combines a large {\it HST}
survey of low$-z$ \ion{O}{6} absorption systems observed on sight
lines to low$-z$ quasars (Tripp et al. 2004) and a ground based survey
to measure the redshifts and properties of the galaxies foreground to
the background QSOs.  The ground based survey is done in two steps:
first, multi-band (U,B,V,R and I) imagery is obtained to identify the
galaxies and to estimate their photometric redshifts. Then,
spectroscopic redshifts are obtained for the galaxies that are
potentially (according to the photometric redhshifts) at lower
redshift that the background object.  As part of the large {\it HST}
survey, we have observed the quasar HS0624+6907 ($z_{\rm QSO}$ =
0.3700) with the E140M echelle mode of the Space Telescope Imaging
Spectrograph (STIS) on board the {\it Hubble Space Telescope}. We have
also obtained multiband images and spectroscopic redshifts of galaxies
in the \hs0624 field.  The sight line to \hs0624 passes by several
foreground Abell clusters (\S~\ref{sec:abell_clusters}) and provides
an opportunity to search for gas in large-scale filaments.  We shall
show that gas (absorption systems) and galaxies are detected at the
redshifts of the structures delineated by the Abell clusters in this
direction.  While the absorbing gas is intergalactic, and it is likely
that we are probing gas in cosmic web filaments, the properties of
these absorbers are surprising.  Instead of low-metallicity WHIM gas,
we predominantly find cool, photoionised, and high-metallicity gas in
these large-scale structures.

This paper is organized as follows. The observations and data
reduction procedures are described in \S2, including {\it HST}/STIS
and {\it Far Ultraviolet Spectroscopic Explorer} (\FUSE) observations as well
as ground-based imaging and galaxy redshift measurements. In \S3, we
present information on the foreground environments probed by the
\hs0624 sight line, derived from the literature on Abell clusters and
from our new galaxy redshift survey.  The absorption-line measurement
methods are described in \S4, and we investigate the physical state
and metallicity of the absorbers in \S5. Section 6 reviews the
properties of the full sample of Ly$\alpha$ lines derived from the
STIS spectrum with emphasis on the search for broad Ly$\alpha$ lines.
Section 7 discusses the implications of this study, and we summarize
our conclusions in \S8. Throughout this paper, we use the following
cosmological parameters: $h_{70}=H_0/70$~\kms, $\Omega_m=0.3$ and
$\Lambda_o=0.7$.

\section{Observations}

\subsection{Ultraviolet QSO Spectroscopy}

\hs0624\ was observed with STIS on 2 Jan. 2002 and 23-24 Feb. 2002 as part of
 a Cycle 10 {\it HST} observing program (ID=9184) .  The echelle
spectrograph was used with the E140M grating which provides a
resolution of 7~\kms\ FWHM and covers the 1150$-$1730~\AA\ range with
only a few small gaps between orders at wavelengths greater than
1630~\AA.  The $0\farcs 2 \times 0\farcs 06$ entrance aperture was
used to minimize the effect of the wings of the line spread function.
The total exposure time was 61.95~ksec.  The data were reduced as
described in Tripp et al.  (2001) using the STIS Team version of
CALSTIS at the Goddard Space Flight Center.  The final signal-to-noise
(S/N) per resolution element is 3 at 1150~\AA, increases
linearly to 14 at 1340\AA\ and then decreases to 7 at
1730~\AA. For further information on the design and performance of
STIS, see Woodgate et al. (1998) and Kimble et al. (1998).

HS0624+6907 was also observed by the \FUSE\ PI Team on several
occasions between 1999 November and 2002 February (Program IDs
P1071001, P1071002, S6011201, and S6011202) .  \FUSE\ records spectra
with four independent spectrographs (``channels''), two with SiC
coatings for coverage of the 905$-$1105 \AA\ wavelength range, and two
with LiF coatings optimized to cover 1100$-$1187 \AA\ (see Moos et al.
2000,2002 for details about \FUSE\ design and performance).  The
spectrograph resolutions range from 20$-$30 km s$^{-1}$ (FWHM).  For
HS0624+6907, the total integration time in the LiF1 channel was 110
ksec; the other channels had somewhat lower integration times due to
channel coalignment problems during some of the observations. We have
retrieved the \FUSE\ spectra from the archive and have reduced the
data using CALFUSE version 2.4.0 as described in Tripp et al.
(2005). Because the spectra in the individual channels have modest S/N
ratios, we have aligned and combined all available LiF channels to
form the final spectra that we used for our measurements (we find that
combining all available LiF data does not degrade the spectral
resolution). For the spectral range uniquely covered by the SiC
channels, we used only the SiC2a data. Finally, we compared absorption
lines of comparable strength (e.g., \ion{Fe}{2} $\lambda$1144.94
vs. \ion{Fe}{2} $\lambda$1608.45) observed by {\it FUSE} and STIS in
order to align the {\it FUSE} spectrum with the STIS spectrum and
thereby correct the wavelength zero point of the \FUSE\ data.

\subsection{Optical Galaxy Imaging and Spectroscopy}\label{sec:optgal}

One of  the primary goals of our  low$-z$ QSO absorption line program is  to study the
connections  between  galaxies   and  absorption  systems.   These
studies require good imaging (for  galaxy target selection and information on
individual galaxies of interest)  followed by optical spectroscopy for
accurate redshift measurements.  To initiate the galaxy-absorber study
toward \hs0624 , we first obtained a $10' \times10'$ mosaic of images
centered on the QSO with  SPICam on the Apache Point Observatory (APO)
3.5m telescope on 2002 October  5. Subsequently, we obtained images of
a larger field  in better seeing with the NOAO 8k$\times$8k CCD
mosaic  camera  (MOSA, Muller  et  al.  1998), on  the  Kitt  Peak  National
Observatory (KPNO) 4m telescope.   The  SPICam images  were  used to  select
targets for  the first spectroscopic observing run,  but thereafter we
only used the better-quality MOSA images.

\begin{table*}
\begin{minipage}{\tableminisize}
\caption{Optical Imaging Observations of \hs0624 \label{tab:obslog}}
\begin{tabular}{@{}lccccc}
\hline
 & \multicolumn{5}{c}{Filters} \\
 & \colhead{\it U} & \colhead{\it B} & \colhead{\it V} & \colhead{\it R} & \colhead{\it I}\\
 \hline
UT Observation Date & 2003 Jan. 30 & 2003 Jan. 30 & 2003 Jan. 29 & 2003 Jan. 29 & 2003
 Jan. 29 \\
 Integration Time (s)& 7$\times$900 & 6$\times$400 & 5$\times$360 & 5$\times$360 & 5$\times$600 \\
\hline
\end{tabular}
\end{minipage}
\end{table*}

\begin{figure*}
\begin{center}
\resizebox{1.0\hsize}{!}{\includegraphics{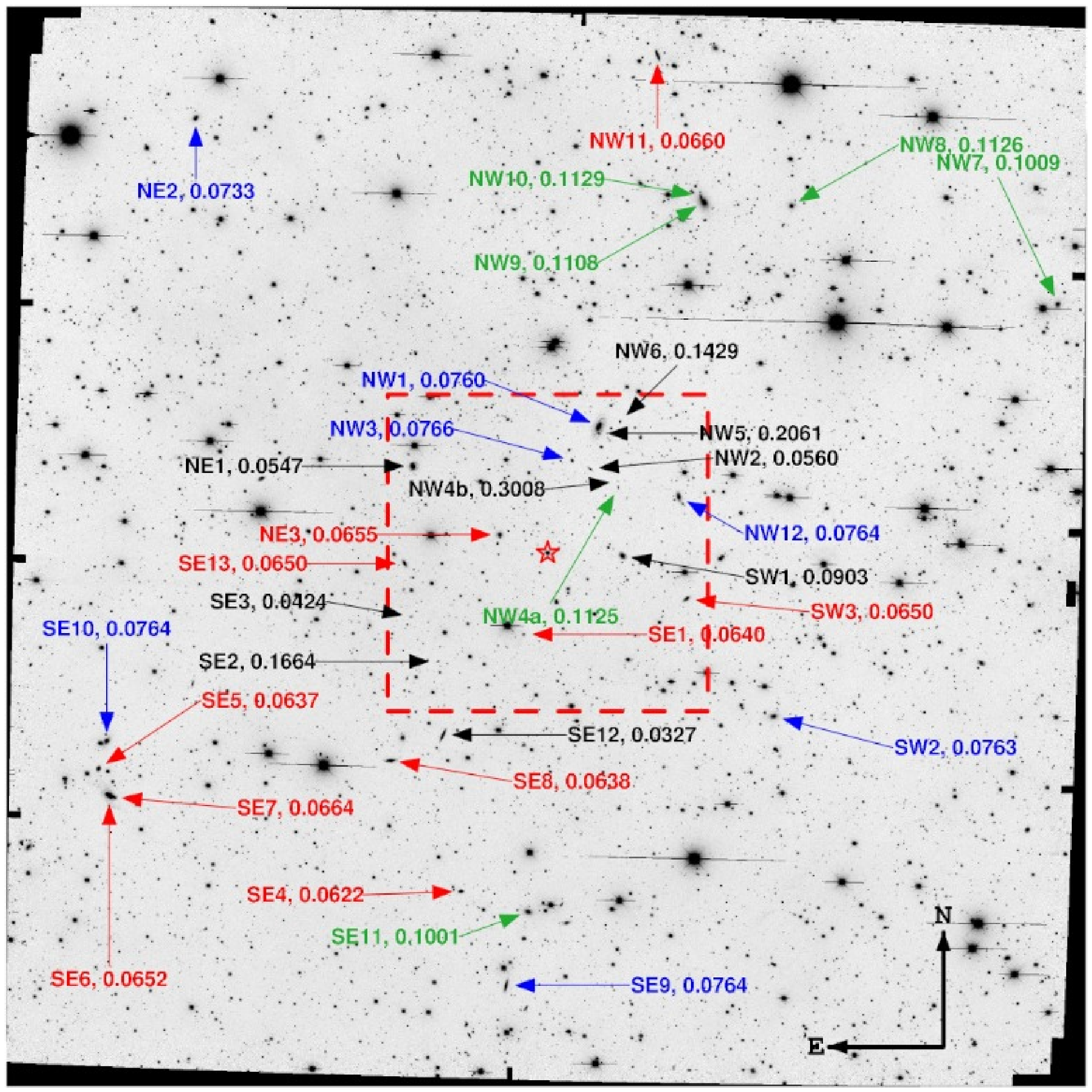}}   
\caption{KPNO 4m {\it R}-band image of the field around
\hs0624. Galaxies are labeled with their ID names (from
Table~\ref{tab:spec_red}) and spectroscopic redshifts when available.
Theses redshift labels are colour-coded as follows: red for
\zbtw{0.062}{0.066} (10 galaxies), blue for \zbtw{0.073}{0.077} (7
galaxies), green for \zbtw{0.100}{0.114} (6 galaxies) and black for
the other redshifts. A close up of the 10'$\times$10' region centered
on the QSO (dashed box) is shown in Figure
\ref{fig:fieldzoom}.\label{fig:field}}
\end{center}
\end{figure*}

\begin{figure*}
\begin{center}
\resizebox{1.0\hsize}{!}{\includegraphics{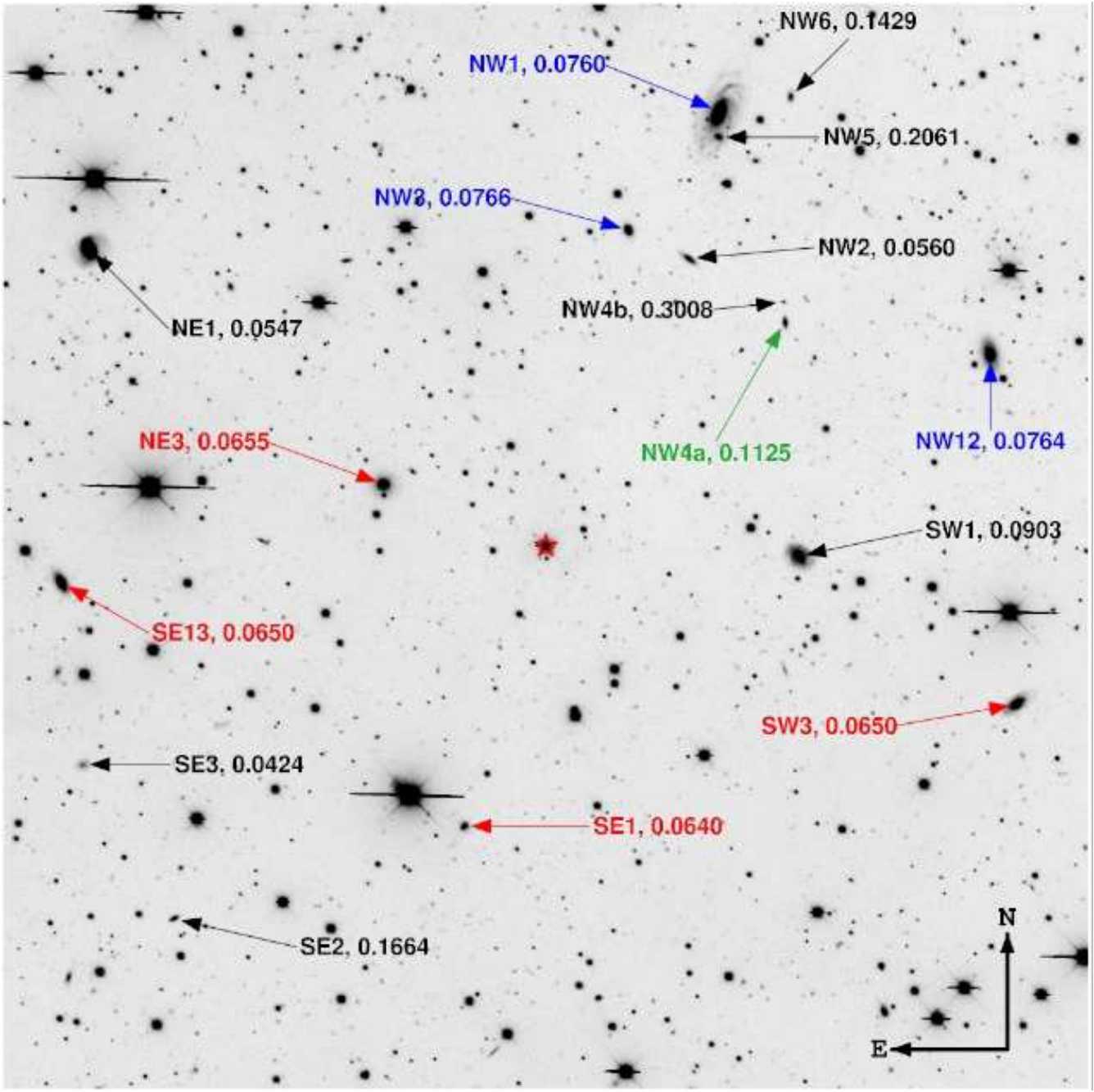}}
\caption{Close-up of the KPNO image shown in Figure
\ref{fig:field}. This portion of the image shows the $10' \times10'$
region centered on the QSO and is labeled and colour-coded as described
in Figure~\ref{fig:field}. At redshift $z\simeq$ 0.065, 0.076 and 0.11, $10'$ corresponds to 750, 860~kpc and 1.2~Mpc respectively.}\label{fig:fieldzoom}
\end{center}
\end{figure*}

\hs0624  was  observed  with  MOSA  on  the  4m on  2003  January
29-30.   The   field   of   view   is  $36'\times36'$   with   a   scale   of
$0\farcs26$/pixel.  As  summarized  in Table~\ref{tab:obslog},  images
were  recorded in  $U, B,  V, R,$  and $I$  with a  standard dithering
pattern  for filling in  gaps between  the CCDs  and for  rejection of
cosmic rays. Photometric standard  stars from Landolt (1992) were also
observed at  regular intervals. During these  observations, the seeing
ranged from $1\farcs$0  to 1$\farcs$3. The data were  reduced with the
IRAF\footnote{IRAF  is distributed by  the National  Optical Astronomy
Observatories, which  are operated by the  Association of Universities
for Research in Astronomy,  Inc., under cooperative agreement with the
National Science Foundation.}  software package {\tt MSCRED} following
standard procedures. The  final R-band MOSA image of \hs0624  is shown
in Figures~\ref{fig:field}~and~\ref{fig:fieldzoom}.

Galaxy  targets  for follow-up  spectroscopy  were  selected from  the
images  using the SExtractor software package  (Bertin \&  Arnouts 1996).  Redshifts  of 29
galaxies were obtained using  the Double Imaging Spectrograph (DIS) on
the APO 3.5~m telescope on the following dates: 2002 November 12, 2003
January 29$-$31,  2003 April  03, 2003 April  21, and 2003  December 25.
Spectra were recorded  using a single 1.5 arcsec  wide slit with total
exposure times ranging  from 360 to 1800 s per  object.  The data were
processed in  the conventional manner, and  were wavelength calibrated
using helium-neon-argon arc-lamp  exposures.  Small zero-point offsets
in wavelength were applied as needed, after comparing observed skyline
wavelengths  with  their  rest  values.  The  spectra  were  typically
recorded at resolutions of $\sim 7-8$~\AA\ FWHM.

%%%%%%%%%%%%%%%%%%%%%%%%%%%%%%%%%%%%%%%%
% Table of the spectroscopic redshifts %
%%%%%%%%%%%%%%%%%%%%%%%%%%%%%%%%%%%%%%%%
{
\def~{\phn}
\begin{table*}
\begin{minipage}{\tableminisize}
\caption{Spectroscopic Redshifts  of Galaxies in the Field of HS 0624+6907\label{tab:spec_red}}
\renewcommand{\footnoterule}{}
\begin{tabular}{llccccccccc}
        &          & \multicolumn{2}{c}{Coordinate (J2000)}         &   & \colhead{$\theta$\tablenotetext{a}{Angular separation to the HS~0624+6907 sightline.}} &
\colhead{$\rho$\tablenotetext{b}{Projected distance to the HS~0624+6907 sightline.} }  &&&&\\
\colhead{\#} & \colhead{ID} & \colhead{RA} & \colhead{DEC} & \colhead{$z$} &\colhead{($'$)} &\colhead{({\hmMpc})} &\colhead{$R$}&\colhead{$V$}&\colhead{$U-V$}&\colhead{$M_R$}
\startdata
   & QSO   & 06:30:02.50 & 69:05:03.99 & 0.3700 & ~0.0 & 0.000 & 13.8\tablenotetext{c}{Due to saturation of the CCD, the magnitudes of the quasar could be underestimated.} & 14.2\tablenotemark{c} &  0.1\tablenotemark{c} & -27.7\tablenotemark{c}\\
01 & SE12  & 06:30:41.70 & 68:58:32.71 & 0.0327 & ~7.4 & 0.290 & 16.6 & 17.8 &  1.0 & -19.2\\
02 & SE3   & 06:30:55.32 & 69:02:41.99 & 0.0424 & ~5.3 & 0.265 & 19.6 & 20.5 &  0.9 & -16.7\\
03 & NE1   & 06:30:56.14 & 69:08:00.90 & 0.0547 & ~5.6 & 0.358 & 15.9 & 17.0 &  0.9 & -21.0\\
04 & NW2   & 06:29:46.66 & 69:08:03.59 & 0.0560 & ~3.3 & 0.216 & 18.6 & 19.5 &  0.9 & -18.4\\
05 & SE4   & 06:30:33.00 & 68:53:02.00 & 0.0622 & 12.3 & 0.887 & 16.8 & 17.9 &  0.9 & -20.5\\
06 & SE5   & 06:32:55.20 & 68:56:59.99 & 0.0637 & 17.4 & 1.282 & 16.8 & 18.3 &  1.3 & -20.5\\
07 & SE8   & 06:31:01.79 & 68:57:35.89 & 0.0638 & ~9.2 & 0.675 & 16.0 & 17.1 &  0.9 & -21.2\\
08 & SE1   & 06:30:11.22 & 69:02:09.61 & 0.0640 & ~3.0 & 0.222 & 18.8 & 19.4 &  0.7 & -18.5\\
09 & SW3   & 06:29:07.80 & 69:03:32.01 & 0.0650 & ~5.1 & 0.384 & 17.4 & 18.3 &  0.9 & -20.0\\
10 & SE13  & 06:30:58.30 & 69:04:34.11 & 0.0650 & ~5.0 & 0.375 & 16.9 & 18.5 &  1.3 & -20.4\\
11 & SE6   & 06:32:50.70 & 68:56:03.00 & 0.0652 & 17.6 & 1.318 & 15.9 & 17.3 &  1.2 & -21.5\\
12 & NE3   & 06:30:21.40 & 69:05:39.70 & 0.0655 & ~1.8 & 0.135 & 16.6 & 18.1 &  1.3 & -20.8\\
13 & NW11  & 06:29:23.48 & 69:22:43.29 & 0.0660 & 18.0 & 1.367 & 16.3 & 17.9 &  1.3 & -21.1\\
14 & SE7   & 06:32:49.20 & 68:56:00.39 & 0.0664 & 17.5 & 1.334 & 17.0 & 18.0 &  0.7 & -20.4\\
15 & NE2   & 06:32:25.55 & 69:20:05.81 & 0.0733 & 19.7 & 1.646 & 16.5 & 18.0 &  1.3 & -21.1\\
16 & NW1   & 06:29:43.65 & 69:09:35.33 & 0.0760 & ~4.8 & 0.417 & 15.9 & 17.3 &  1.1 & -21.8\\
17 & SW2   & 06:28:33.03 & 68:59:26.30 & 0.0763 & ~9.8 & 0.849 & 16.6 & 18.3 &  1.3 & -21.1\\
18 & SE9   & 06:30:14.81 & 68:49:44.79 & 0.0764 & 15.4 & 1.334 & 16.4 & 18.0 &  1.3 & -21.3\\
19 & NW12  & 06:29:11.59 & 69:07:07.89 & 0.0764 & ~5.0 & 0.433 & 16.4 & 18.0 &  1.4 & -21.3\\
20 & SE10  & 06:32:52.40 & 68:57:59.01 & 0.0764 & 16.8 & 1.457 & 16.5 & 18.1 &  1.3 & -21.2\\
21 & NW3   & 06:29:53.77 & 69:08:20.51 & 0.0766 & ~3.4 & 0.293 & 18.3 & 19.6 &  1.1 & -19.4\\
22 & SW1   & 06:29:33.24 & 69:05:01.00 & 0.0903 & ~2.6 & 0.264 & 17.1 & 18.2 &  0.9 & -21.0\\
23 & SE11  & 06:30:06.84 & 68:52:22.20 & 0.1001 & 12.7 & 1.407 & 16.5 & 18.1 &  1.2 & -21.9\\
24 & NW7   & 06:26:43.70 & 69:14:06.91 & 0.1009 & 19.9 & 2.215 & 16.7 & 18.2 &  1.4 & -21.6\\
25 & NW9   & 06:29:03.89 & 69:17:33.40 & 0.1108 & 13.5 & 1.639 & 16.4 & 18.2 &  1.3 & -22.2\\
26 & NW4a  & 06:29:35.43 & 69:07:25.80 & 0.1125 & ~3.4 & 0.415 & 18.5 & 20.2 &  1.2 & -20.1\\
27 & NW8   & 06:28:29.39 & 69:17:28.89 & 0.1126 & 14.9 & 1.832 & 16.7 & 18.5 &  1.3 & -21.9\\
28 & NW10  & 06:29:05.25 & 69:17:46.69 & 0.1129 & 13.7 & 1.685 & 18.1 & 19.9 &  1.3 & -20.5\\
29 & NW6   & 06:29:35.30 & 69:09:44.99 & 0.1429 & ~5.3 & 0.794 & 19.4 & 20.4 &  0.8 & -19.8\\
30 & SE2   & 06:30:44.35 & 69:01:08.09 & 0.1664 & ~5.4 & 0.927 & 19.3 & 20.8 &  1.0 & -20.2\\
31 & NW5   & 06:29:43.52 & 69:09:19.01 & 0.2061 & ~4.6 & 0.927 & 18.4 & 20.3 &  1.4 & -21.7\\
32 & NW4b  & 06:29:35.62 & 69:07:37.91 & 0.3008 & ~3.5 & 0.940 & 21.0 & 22.2 &  0.4 & -19.9
\enddata
%\tablenotetext{a}{Angular separation to the HS~0624+6907 sightline.}
%\tablenotetext{b}{Projected distance to the HS~0624+6907 sightline.} 
%\tablenotetext{c}{Due to saturation of the CCD, the magnitudes of the quasar could be underestimated.}
\end{tabular}
\vspace{-0.6cm}
\end{minipage}
\end{table*}
}

%\clearpage

The redshift measurements were made following the procedure
described by Jenkins~et~al. (2003). We  used the IRAF routine FXCOR to
cross-correlate the  galaxy spectra with  that of the  radial velocity
standard HD~182572.  In general we only used the blue channel DIS data
for  the  cross-correlation, where  the  4000~\AA\  break and  stellar
absorption  lines were most  apparent. Red  channel data  were usually
used to  identify and measure  the wavelengths of  redshifted emission
lines  ([O~{\sc iii}],  H$\beta$, H$\alpha$,  etc.) when  present. The
galaxy   redshifts   obtained   in   this  way   are   summarized   in
Table~\ref{tab:spec_red} and are accurate to between 70 and 170~\kms (which corresponds to a sight line distance displacement uncertainty of 1.0 to 2.4~Mpc for an unperturbed Hubble flow).

We also observed three galaxies with the Echellette Spectrometer and
Imager (ESI; Sheinis et al. 2001) on the 10m Keck II telescope on the
nights of 2004 September 10 and 11 during morning twilight.  We
observed galaxy NE3 (see Table~\ref{tab:spec_red}) in echellette mode
with the 0.5$''$ slit which provides $\approx 30 $~\kms\ spectral
resolution (FWHM).  The fainter SE13 and SW3 galaxies were observed in
low dispersion mode using a 1$''$ slit which affords $R \sim 2000$ at
$\lambda = 5000$~\AA.  The exposures were flat fielded and wavelength
calibrated with the ESIRedux package (Prochaska et al.  2003%
\footnote{http://www.ucolick.org/$\sim$xavier/ESI/index.html}).  The
NE3 redshift was derived from the centroids of the high-resolution
\ion{Na}{1} and H$\alpha$ absorption lines, and the redshift
uncertainty is $\sim$30 km s$^{-1}$.  For SE13 and SW3, redshifts were
measured by fitting \ion{Na}{1}, H$\beta$, and \ion{Ca}{2} H and K,
and the uncertainties are $\sim$150 km s$^{-1}$.

\begin{figure}
%\begin{center}
\resizebox{1.0\hsize}{!}{\includegraphics{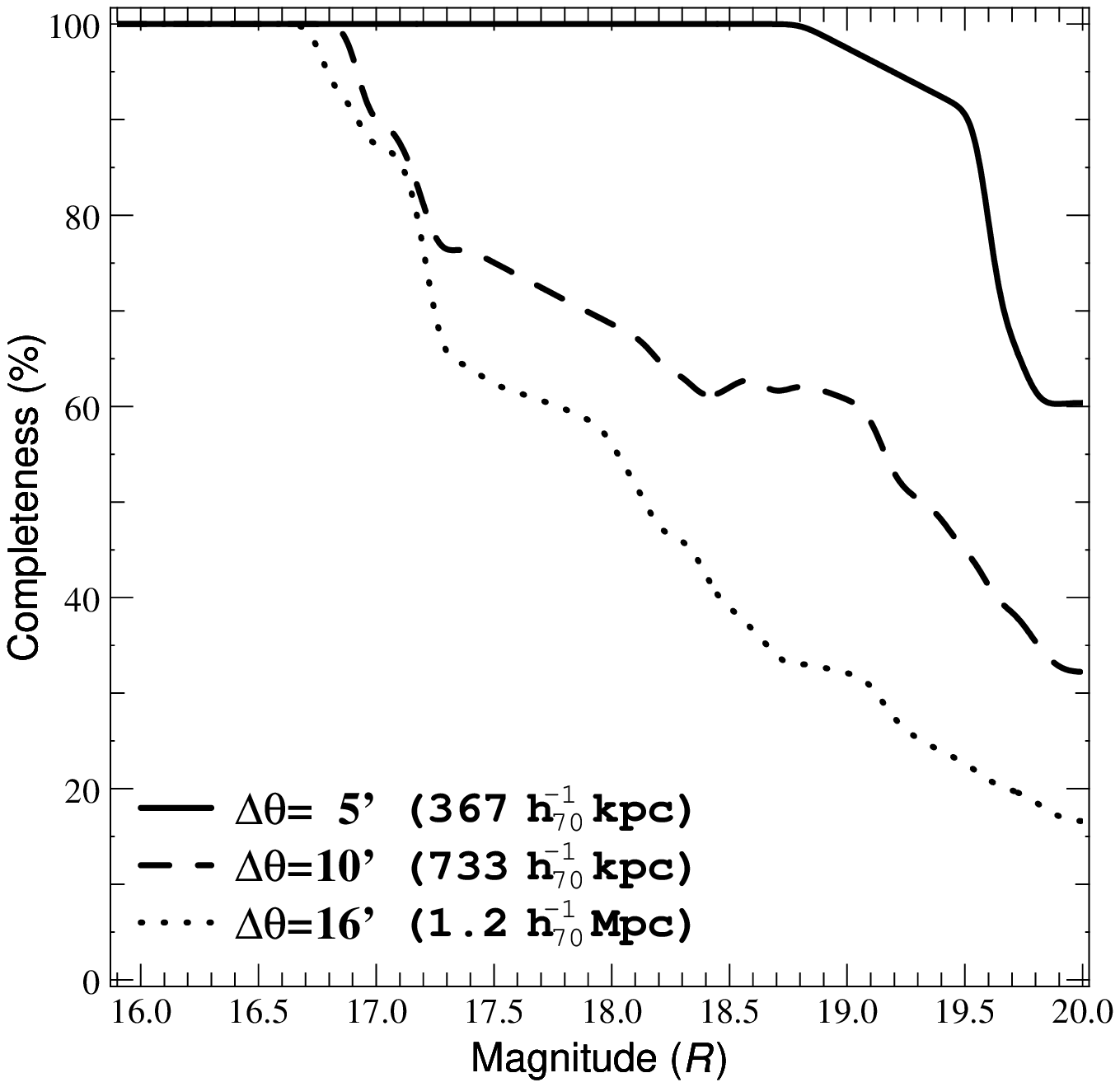}}
\caption{Completeness    of   the   spectroscopic    galaxy   redshift
measurements as a function of  limiting $R$ magnitude for targets within
an angular separation $\Delta \theta$  = 5, 10, and 16$'$ as indicated
in the key.  The key also shows in  parentheses the projected physical
distances  corresponding  to these  $\Delta  \theta$  values  at $z  =
0.0635$.\label{fig:completeness}}
%\end{center}
\end{figure}

The completeness  of our galaxy redshift survey  (i.e., the percentage
of targets  brighter than a  given magnitude in the  SExtractor galaxy
catalog with  good spectroscopic redshifts)  is graphically summarized
in  Figure~\ref{fig:completeness}  as   a  function  of  limiting  $R$
magnitude  and  angular  separation   from  the  sight  line  ($\Delta
\theta$). In the  $10'\times10'$ region centered on \hs0624  , we have
measured spectroscopic redshifts for  all galaxies brighter than $R$ =
19.0, and  the survey is $\approx 60$ per cent  complete for $R <  20$. As we
shall see,  there is  a prominent cluster  of absorption lines  in the
\hs0624  spectrum  at  $z  \approx$  0.0635; at  this  redshift,  5$'$
corresponds to a projected distance  of 367 $h_{70}^{-1}$ kpc, and $R$
= 19.0 corresponds  to $M=-18.3$ or $L = 0.1  L*$ (taking $M_{R}^{*} =
-21.0$  from  Lin  et  al.  1996).  For comparison, the Large Magellanic cloud has a magnitude equal to $M_R=-17$ or $L=0.02 L*$. At this  redshift,  we  have  good
completeness even for low luminosity galaxies.  At larger radii, a substantial
number  of bright  galaxies  are  found, and  our  redshift survey  is
shallower. Nevertheless, within  a 10' radius circle, our  survey is still
60 per cent complete for galaxies brighter than $R < 19$.

\section{Absorber Environment}\label{sec:environment}

Using information gleaned from  the literature in combination with our
galaxy redshift survey, we can identify several large-scale structures
that are pierced by the \hs0624 sight line. In this section we comment
on    these    structures     including    nearby    Abell    clusters
(\S\ref{sec:abell_clusters})  as  well as  smaller  (and closer)  galaxy
groups (\S\ref{sec:close_groups}).

\subsection{Nearby Abell Clusters and Large-Scale Structure}\label{sec:abell_clusters}

Clusters are clustered and often reveal even larger cosmic structures,
i.e., superclusters  (Einasto et al. 2001 and  references therein). In
cosmological  simulations,  clusters  are  found at  the  nodes  where
large-scale filaments  connect. To  test the fidelity  of cosmological
simulations,  which  are   now  being  used  in  a   wide  variety  of
astrophysical analyses,  it is  important to search  for observational
evidence  of  the expected  {\it gaseous} filaments  feeding  into  clusters and  to
measure the  properties of  the filaments. The  sight line  to \hs0624
passes through a  region of relatively high Abell  cluster density and
is well-suited for investigation of this topic.

%%%%%%%%%%%%%%%%%%%%%%%%%%%%%%%%%%%%%%%%%%%
% Map of the Abell clusters around HS0624 %
%%%%%%%%%%%%%%%%%%%%%%%%%%%%%%%%%%%%%%%%%%%
\begin{figure}
\begin{center}
\resizebox{1.0\hsize}{!}{\includegraphics{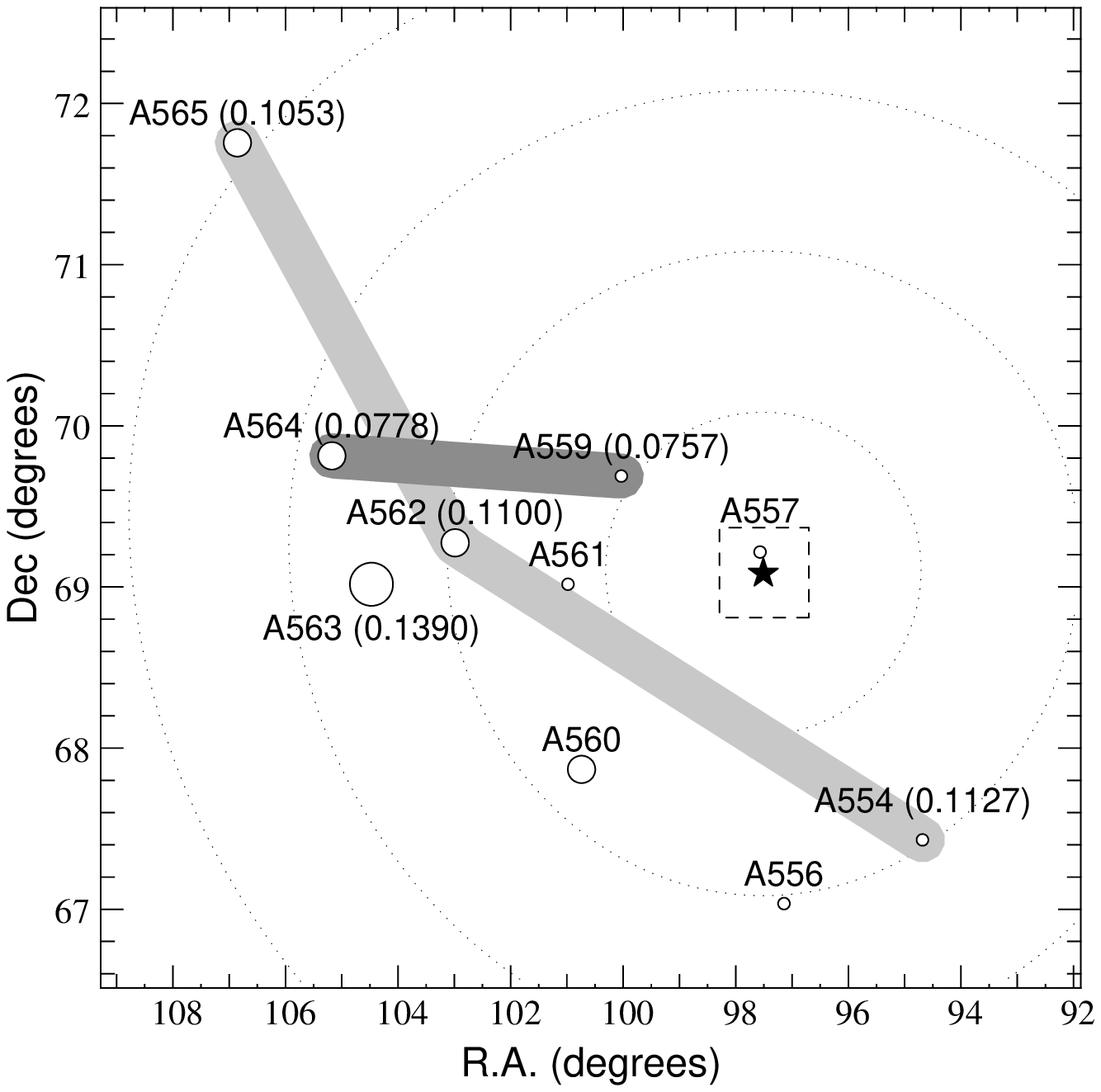}}
\caption{Position  of the  Abell  clusters around  the  sight line  of
HS0624$+$6907.   The  size  of  the  circles  indicates  the  relative
population richness  of the  cluster.  If known,  the redshift  of the
Abell cluster is indicated in parentheses. The shaded regions schematically indicate
the possible  large scale structures  traced by the Abell  clusters at
\zapp{\zvb} and \zapp{\zvc}.  The  dashed rectangle corresponds to the
limits of the  KPNO image and the star to the  position of the quasar.
The 1,  2, 3 and 4 degrees angular  separation to the quasar  are shown by
the dotted circles. At redshift z$\simeq$ 0.076 and 0.110, 2 degrees correspond to 10 and 14~Mpc respectively.}\label{fig:abell_map}
\end{center}
\end{figure}

\fig{fig:abell_map} shows  the positions of  Abell clusters around
the sight  line to \hs0624, including the
cluster richness class and spectroscopic redshift (when available from
the  literature). The  density of  Abell  clusters in  this region  is
relatively high compared to the  vicinity of the other clusters in the
Abell  catalog:  the  number  of  Abell clusters  within  2$^{\rm  o}$
(3$^{\rm o}$) of \abell{557} (the cluster closest to the \hs0624 sight
line) is 2  (1.6) times larger than the  average number within 2$^{\rm
  o}$ (3$^{\rm o}$) of all Abell clusters.  Einasto et al. (2001) have
identified  two  superclusters in  the  direction  of \hs0624.   Their
supercluster    SCL71     (at    $z    \approx     0.110$)    includes
\abell{554},\abell{557},  \abell{561},  \abell{562},  and  \abell{565}
while  SCL72   (at  $z   \approx  0.077$)  includes   \abell{559}  and
\abell{564}.  However,   \abell{557}  and   \abell{561}  do   not  have
spectroscopic redshifts,  and based on our  spectroscopic redshifts in
the field  of HS0624+6907/Abell557 (see  Table~\ref{tab:spec_red}), it
appears  likely  that the  visually  identified  \abell{557}  is a  false
cluster due to the superposition of galaxy groups at several different
redshifts.    To  be   conservative,   we  only   use  clusters   with
spectroscopic redshifts to  identify large-scale structures. The
clusters at \zapp{\zvb} (\abell{564} and \abell{559}) are separated by
4.7~\hmMpc\ from each other, and the clusters at \zapp{\zvc} ( \abell{565}, \abell{562}
and  \abell{554})  are  separated  by  3.9 and 8.7~\hmMpc.   According  to
Colberg et al. (2004), in  cosmological simulations, more than 85 per cent of
the  clusters with a  separation lower  than 10~\hmMpc\  are connected
with  a filament.   We  will  show in  subsequent  sections that  both
absorption lines in the spectrum  of \hs0624 and galaxies close to the
sight  line are  detected  at the  redshifts  of both  of these  Abell
cluster  structures,  which  indicates  that gaseous  filaments
connect into the clusters.

\subsection{Individual Galaxies and Groups}\label{sec:close_groups}

\begin{figure}
\begin{center}
\resizebox{1.0\hsize}{!}{\includegraphics{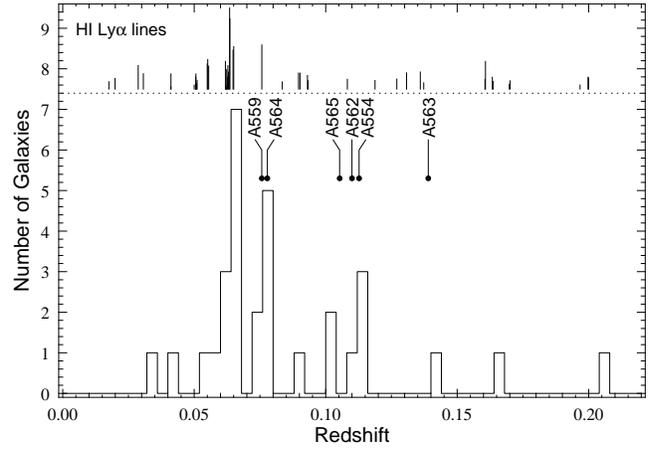}}
\caption{Histogram  of galaxy redshifts  in the  field of  \hs0624. \ion{H}{1}
Lyman-$\alpha$ systems are  indicated on the top of  the figure with a
line height proportional to  the absorption line rest equivalent width. Also
reported   on  the   figure are  the   known  redshifts   of   the  Abell
clusters shown in Figure \ref{fig:abell_map}. }\label{fig:zgaldist}
\end{center}
\end{figure}

\begin{figure}
\begin{center}
\resizebox{1.0\hsize}{!}{\includegraphics{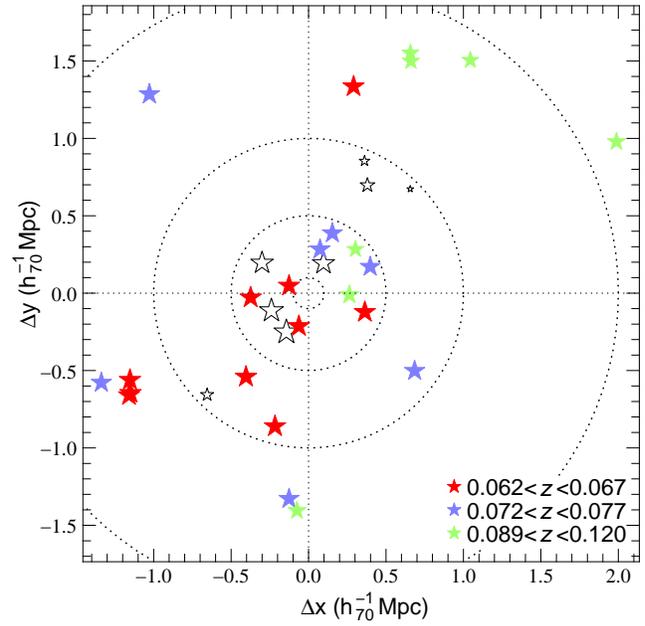}}
\caption{Map of the galaxy distribution around the quasar \hs0624\ in
projected physical coordinates (north is up and east is to the
left). The QSO is at (x,y) = (0,0), and the galaxies with
spectroscopic redshifts (see Table~\ref{tab:spec_red}) are shown with
stars. The symbol size is inversely proportional to redshift. Some 
galaxies are colour-coded by redshift in ranges of $z$ which correspond to 
peaks in the redshift distribution shown in Figure~\ref{fig:zgaldist} or the redshifts of nearby Abell clusters (Figure~\ref{fig:abell_map}).  Distances of 0.1,
0.5, 1.0 and 2.0~\hmMpc\ to the quasar are indicated by dotted
circles. }\label{fig:gal_colour}
\end{center}
\end{figure}

In this section we offer
some brief comments  about specific galaxies and galaxy groups close to the sight line of \hs0624, as   revealed by our optical spectroscopy.
%\fig{fig:field}  shows the  36'~$\times$~36' MOSA  image of  the field
%around  the  sight  line  labeled with  spectroscopic  redshifts  from
%Table~\ref{tab:spec_red}.  The   quasar  is  in  the   center  and  is
%identified  by  the  red  star;  an  expanded  image  of  the  central
%10'$\times$10' region  is provided in \fig{fig:fieldzoom}. 
We place these observations in the context of the Abell clusters 
described in the previous section. We plot in
\fig{fig:zgaldist}  the  redshift distribution  of  the galaxies  from
Table~\ref{tab:spec_red}.  From  this  figure  we can  identify  three
galaxy groups:  two galaxy  groups appear to  be present  at redshifts
similar  to those  of  the  Abell clusters,  i.e,  at $z\sim0.077$  (7
galaxies)   and  $z\sim0.11$   (6  galaxies).   This   indicates  that
%large-scale cosmic filaments do  indeed connect to these Abell cluster
%structures.
the filament of galaxies connecting \abell{559} to \abell{564} must
extend more then 3 degrees (15~\hmMpc) west from \abell{559} and
structure linking \abell{562} and \abell{554} likely extends by at least
3 degrees (22~\hmMpc) in order to cross the QSO sight line.  However,
the most prominent group in \fig{fig:zgaldist} includes 10 galaxies at
\zapp{\zva}, which does not match up with any Abell cluster with a
known spectroscopic redshift. To show the spatial distribution of
galaxies in the three prominent groups in the MOSA field,
\fig{fig:gal_colour} provides a colour-coded map of projected spatial
coordinates of the galaxies. We see that the group at $z \approx
0.064$ is mostly southeast of the sight line while the galaxies
associated with Abell 554/562/565 at \zapp{0.11} are predominantly
northwest of \hs0624 . The galaxies associated with the Abell 559/564
\zapp{0.07} supercluster appear to extend from the southwest across
the sight line to the northeast.

We  note  that spectroscopic  redshifts  are  not
available   for    several   of    the   clusters   shown    in   Fig.
\ref{fig:abell_map}, including  the one that  is closest to  the sight
line,  \abell{557}. However,  Abell  clusters  are visually  identified
without  the   aid  of   spectroscopy,  and  it   can  be   seen  from
Figures~\ref{fig:field} and  \ref{fig:gal_colour} that several discrete
groups are  found at  the location  of \abell{557}.   It is  likely that
\abell{557}  is not a  true cluster but  rather is the  superposition of
several groups in projection.

Is gas also present in these large-scale cosmic filaments? To address
this question, we searched the spectrum of \hs0624 for any absorption
counterparts at the redshifts of galaxies and galaxy structures near
the QSO sight line.   The redshifts of the  \ion{H}{1} \lya\ systems  that we have
identified and measured (see  \S\ref{sec:absline}) in the spectrum of
\hs0624 are  plotted at the  top of \fig{fig:zgaldist}; the  length of
the line is proportional to the rest equivalent width of the  \lya\ line. This  search has
revealed three interesting results:  First, when a galaxy is located at
an impact parameter  $\rho  \lesssim 500 $~\hmkpc\ from  the sight
line, \lya\  absorption is  almost always found  within a  few hundred
\kms\  of  the galaxy  redshift  (compare Table~\ref{tab:spec_red}  to
Table~\ref{tab:lyalist}),  consistent with  the  findings of  previous
studies  (e.g., Lanzetta  et al.  1995; Tripp  et al.  1998;  Impey et
al.  1999;   Chen  et  al.  2001;   Bowen  et  al.   2002;  Penton  et
al. 2002). Second, strong absorption is clearly detected at the
redshift of the $z\sim0.07$ Abell  564/559 supercluster. This absorption system is
detected in \lya , \lyb , and the \ion{C}{4} $\lambda \lambda$
1548.20, 1550.78 doublet (\S\ref{ss:z076}), and the absorption
redshift is very similar to that of \abell{559}.  Weak \lya\
absorption is also detected at $z_{\rm abs}$ = 0.10822, which is
within 500 \kms\ of the Abell 554/562/565 filament. Evidently, and not
surprisingly, gas is also found in the filaments that feed into the
clusters near \hs0624. Third, Figure~\ref{fig:zgaldist} qualitatively
indicates that the strongest \lya\ lines are situated in the regions
of highest galaxy density, which is similar to the conclusions of
Bowen et al. (2002) and C{\^ o}t\'e et al. (2005).

Could these \lya\ absorbers simply arise in the halos of individual
galaxies? As we show in the next section, the strong \ion{H}{1} system
at \zabsapp{\zva} is comprised of a large number of components spread
over 1000 \kms.  Such kinematics are unprecedented in single galaxy
halos. In addition, we find no obvious pattern that shows a connection
between individual \lya\ lines and individual galaxies in this
complex. In the case of this strong \ion{H}{1} system at
\zabsapp{\zva}, the closest observed galaxy to the sight line around
this redshift has $\rho = 135\,h_{70}^{-1}{\rm kpc} $ (NE3 in table
\ref{tab:spec_red}). However, a closer and fainter galaxy could have
been missed by the spectroscopic survey.  Using photometric redshifts
(measured as described in Chen et al. 2003) to cull the distant
background galaxies with photometric redshifts $> 0.2$, we find that
there are only four galaxies closer to the sight line than NE3 that
could be near \zapp{0.064}.  These four objects are only candidates
since photometric redshifts have substantial uncertainties. However,
if we assume the redshift of these candidates to be $z=0.064$, the
closest one to \hs0624 has still a large impact parameter $\rho =
85\,h_{70}^{-1}$~kpc. In the case of \lya\ at $z_{\rm abs}$ = 0.07573,
the closest galaxy in projection is NW3 at $\rho =
293\,h_{70}^{-1}$~kpc.  It is conceivable that this absorption
originates in the large halo of this particular galaxy, but we note
that three galaxies are found at $\rho \leq 500\,h_{70}^{-1}$~kpc at
this $z$, and many other origins are possible (e.g., intragroup gas or
tidally stripped debris). Finally, the absorption at $z_{\rm abs} =
0.10822$ is at a substantial distance (415~\hmkpc) from the nearest
known galaxy (NW4a) and is unlikely to be halo gas associated with
that object.

\section{Absorption Line Measurements}\label{sec:absline}
We now turn  to the absorption-line measurements. As  discussed in the
previous section, Figure~\ref{fig:zgaldist} compares the distributions
of galaxies and Ly$\alpha$ lines  in the direction of \hs0624.  We
have  measured the  column  densities and  Doppler  parameters of  the
Ly$\alpha$ lines  in the spectrum  of \hs0624 using the  Voigt profile
decomposition   software    VPFIT   (see   Webb   1987\footnote{http://www.ast.cam.ac.uk/$\sim$rfc/vpfit.html}). Table~\ref{tab:lyalist}
summarizes  the \ion{H}{1}  equivalent widths,  column  densities, and
$b-$values measured  in this fashion  (some of the lines  are strongly
saturated  and consequently  Voigt  profile fitting  does not  provide
reliable measurements; we discuss our treatment of these cases below).

\begin{figure}
\begin{center}
\resizebox{1.0\hsize}{!}{\rotatebox{-90}{\includegraphics{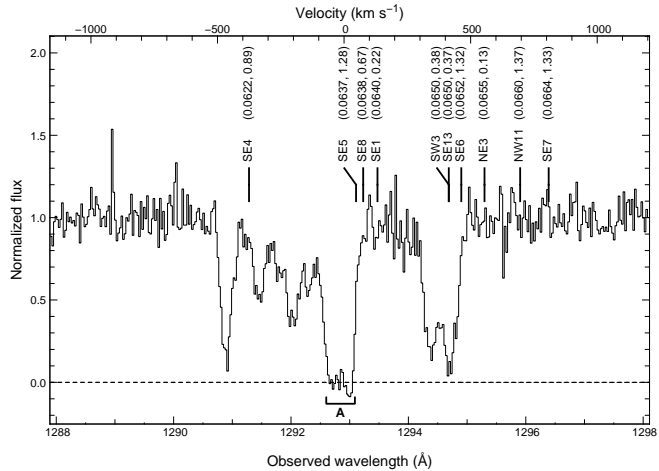}}}
\caption{STIS spectrum of \hs0624 between 1288~\AA\ and 1298~\AA. All
the absorption lines shown are due to a \ion{H}{1} \lya\ complex
covering a velocity range $\simeq$1000~km~s$^{-1}$. The velocity of
the upper axis is relative to \lya\ at $z=0.06352$. 
The tick marks show the galaxy redshifts from our
spectroscopic survey (\S\ref{sec:optgal}).  Apart from the
identification name, the labels indicate the redshifts of
the galaxies and their projected distances (in \hmMpc) from the QSO sight
line, respectively.\label{fig:hi00635}}
\end{center}
\end{figure}

A particularly dramatic cluster of Ly$\alpha$ lines is evident at
$z_{\rm abs} \approx$ 0.0635 in Figure~\ref{fig:zgaldist}.  The
portion of the STIS spectrum of HS0624+6907 covering this Ly$\alpha$
cluster is shown in Figure~\ref{fig:hi00635}.  To avoid confusion with
galaxy clusters, we will hereafter refer to this group of \lya\ lines
as a \lya\ ``complex''. This complex contains at least 13 \ion{H}{1}
components spread over a velocity range of 1000 km s$^{-1}$.  We will
refer to the strongest Ly$\alpha$ absorption in
Figure~\ref{fig:hi00635} at $z_{\rm abs}$ = 0.06352 as ``component
A''.  Component A is detected in absorption in the \ion{H}{1}
Ly$\alpha$, Ly$\beta$, and Ly$\gamma$ transitions as well as the
\ion{Si}{3} $\lambda$1206.50, \ion{Si}{4} $\lambda \lambda$1393.76,
1402.77, and \ion{C}{4} $\lambda \lambda$1548.20, 1550.78 lines.  Low
ionisation stages such as \ion{O}{1}, \ion{C}{2}, and \ion{Si}{2} are
not detected at the redshift of Component A or at the redshifts of any
of the other components evident in Figure~\ref{fig:hi00635}.  The
\ion{O}{6} doublet at the redshifts of the Ly$\alpha$ cluster in falls
in a region that is partially blocked by Galactic H$_{2}$ and
\ion{Fe}{2} lines.  Nevertheless, much of the region is free from
blending, and we find no evidence for \ion{O}{6} absorption.  We also
do not see the \ion{N}{5} doublet.

\begin{table*}
\begin{minipage}{\tableminisize}
\caption{Equivalent Widths, Column Densities, and Doppler Parameters of \ion{H}{1} Ly$\alpha$ Lines in the Spectrum of HS 0624$+$6907\label{tab:lyalist}}
\begin{tabular}{lccl|lccl}
 & $W_{obs}$ & $\log({\rm N})$ & $b$ & & $W_{obs}$  & $\log({\rm N})$ & $b$ \\
$z$ & (m\AA)  & (\cmmd) & (\kms) & $z$ & (m\AA)  & (\cmmd) & (\kms)
\startdata
0.017553$\pm$~1.0e-5     & ~45$\pm$10 & 12.96$\pm$0.05 & ~29$\pm$~4.3     &0.207540$\pm$~0.5e-5     & 150$\pm$~9 & 13.48$\pm$0.02 & ~27$\pm$~1.5     \\
0.030651$\pm$~0.4e-5     & ~99$\pm$~9 & 13.36$\pm$0.03 & ~22$\pm$~1.7     &0.213232$\pm$~1.6e-5     & ~98$\pm$14 & 13.22$\pm$0.05 & ~45$\pm$~5.6     \\
0.041156$\pm$~0.8e-5     & 104$\pm$11 & 13.33$\pm$0.03 & ~41$\pm$~3.0     &0.219900$\pm$~2.3e-5     & 143$\pm$15 & 13.39$\pm$0.05 & ~60$\pm$~8.6     \\
0.053942$\pm$~0.6e-5     & ~85$\pm$~7 & 13.26$\pm$0.04 & ~24$\pm$~2.3     &0.223290$\pm$~0.3e-5     & 256$\pm$12 & 13.86$\pm$0.02 & ~25$\pm$~0.9     \\
0.054367$\pm$~4.1e-5$^{a}$ & ~65$\pm$13 & 13.09$\pm$0.11 & ~60$\pm$19.2$^b$ &0.232305$\pm$~2.8e-5$^a$ & 125$\pm$13 & 13.33$\pm$0.08 & ~44$\pm$~7.7$^b$ \\
0.054829$^{a,c}$ & 458$\pm$10 & $\sim$14.5$^c$ & ~$\sim$35$^c$    &0.232547$\pm$~2.3e-5$^a$ & ~44$\pm$10 & 12.86$\pm$0.21 & ~24$\pm$~7.3     \\
0.055153$\pm$~7.8e-5$^a$ & 237$\pm$14 & 13.68$\pm$0.17 & ~84$\pm$30.7$^b$ &0.240599$\pm$~0.6e-5     & 110$\pm$10 & 13.33$\pm$0.04 & ~20$\pm$~2.0     \\
0.061879$\pm$~0.4e-5     & 184$\pm$~6 & 13.77$\pm$0.03 & ~21$\pm$~1.4     &0.252251$\pm$~1.2e-5     & ~55$\pm$11 & 12.96$\pm$0.06 & ~24$\pm$~4.2     \\
0.062014$\pm$~1.0e-5     & ~21$\pm$~4 & 12.63$\pm$0.17 & ~~8$\pm$~4.7     &0.268559$\pm$~2.1e-5     & ~68$\pm$14 & 13.03$\pm$0.05 & ~51$\pm$~7.2     \\
0.062150$\pm$~1.5e-5$^a$ & ~13$\pm$~4 & 12.41$\pm$0.22 & ~10$\pm$~7.9     &0.272240$\pm$~0.6e-5     & ~37$\pm$~8 & 12.80$\pm$0.06 & ~12$\pm$~2.2     \\
0.062343$\pm$~0.8e-5$^a$ & 128$\pm$~7 & 13.45$\pm$0.05 & ~30$\pm$~4.0     &0.279771$\pm$~1.7e-5$^a$ & 174$\pm$13 & 13.50$\pm$0.06 & ~34$\pm$~4.9     \\
0.062647$\pm$~2.5e-5$^a$ & 101$\pm$~7 & 13.31$\pm$0.14 & ~35$\pm$12.3     &0.280171$\pm$~0.7e-5$^a$ & 576$\pm$15 & 14.32$\pm$0.02 & ~43$\pm$~1.9$^b$ \\
0.062762$\pm$~0.7e-5$^a$ & ~39$\pm$~3 & 12.95$\pm$0.28 & ~~8$\pm$~3.7     &0.295307$\pm$~0.7e-5     & 309$\pm$15 & 13.80$\pm$0.02 & ~42$\pm$~2.0     \\
0.062850$\pm$~1.2e-5$^a$ & 110$\pm$~5 & 13.42$\pm$0.14 & ~20$\pm$~7.0     &0.296607$\pm$~0.9e-5     & 203$\pm$18 & 13.54$\pm$0.02 & ~52$\pm$~2.9     \\
0.063037$\pm$~1.4e-5$^a$ & 101$\pm$~6 & 13.33$\pm$0.13 & ~27$\pm$~8.8     &0.308991$\pm$~0.6e-5     & 167$\pm$12 & 13.49$\pm$0.03 & ~28$\pm$~1.8     \\
0.063456$\pm$~1.6e-5$^a$ & 569$\pm$~9 & 14.46$\pm$0.30 & ~48$\pm$~8.4$^b$ &0.309909$\pm$~5.5e-5$^a$ & 246$\pm$18 & 13.61$\pm$0.10 & ~66$\pm$12.3$^b$ \\
0.063481$\pm$~1.6e-5$^a$ & 443$\pm$~6 & 15.27$\pm$0.13 & ~24$\pm$~5.5     &0.310454$\pm$~8.0e-5$^a$ & 170$\pm$17 & 13.43$\pm$0.33 & ~62$\pm$40.3$^b$ \\
0.063620$\pm$~2.7e-5$^a$ & 153$\pm$~4 & 14.29$\pm$0.38 & ~10$\pm$~5.6     &0.310881$\pm$14.4e-5     & ~88$\pm$16 & 13.13$\pm$0.43 & ~51$\pm$27.7     \\
0.064753$\pm$~0.9e-5$^a$ & 257$\pm$~9 & 13.87$\pm$0.04 & ~33$\pm$~3.0     &0.312802$\pm$~4.4e-5     & 257$\pm$18 & 13.65$\pm$0.10 & ~54$\pm$~9.3     \\
0.065016$\pm$~0.8e-5$^a$ & 282$\pm$~8 & 13.97$\pm$0.04 & ~31$\pm$~2.7     &0.313028$\pm$~1.6e-5     & ~72$\pm$10 & 13.09$\pm$0.24 & ~17$\pm$~6.8     \\
0.075731$\pm$~0.2e-5     & 292$\pm$~8 & 14.18$\pm$0.03 & ~24$\pm$~0.8     &0.313261$\pm$~4.7e-5     & 244$\pm$19 & 13.62$\pm$0.10 & ~55$\pm$10.9     \\
0.090228$\pm$~4.2e-5     & 106$\pm$12 & 13.29$\pm$0.08 & ~76$\pm$13.7     &0.317901$\pm$~1.2e-5     & 139$\pm$18 & 13.37$\pm$0.04 & ~34$\pm$~3.6     \\
0.130757$\pm$~1.0e-5     & 114$\pm$~9 & 13.34$\pm$0.04 & ~34$\pm$~3.6     &0.320889$\pm$~0.4e-5     & 349$\pm$14 & 13.97$\pm$0.02 & ~31$\pm$~1.2     \\
0.135966$\pm$~3.9e-5     & 119$\pm$11 & 13.33$\pm$0.10 & ~57$\pm$10.7     &0.327245$\pm$~5.0e-5$^a$ & 316$\pm$21 & 13.73$\pm$0.32 & ~69$\pm$15.6$^b$ \\
0.160541$\pm$~5.0e-5$^a$ & ~69$\pm$~7 & 13.08$\pm$0.21 & ~34$\pm$10.3     &0.327721$\pm$38.7e-5$^a$ & 264$\pm$26 & 13.61$\pm$0.43 & 115$\pm$62.1$^b$ \\
0.160744$\pm$~1.0e-5$^a$ & 200$\pm$~7 & 13.66$\pm$0.05 & ~30$\pm$~2.4     &0.332674$\pm$~1.1e-5     & 202$\pm$18 & 13.55$\pm$0.04 & ~38$\pm$~3.4     \\
0.199750$\pm$~0.6e-5     & ~87$\pm$~9 & 13.24$\pm$0.05 & ~17$\pm$~2.0     &0.339759$\pm$~0.3e-5     & 647$\pm$13 & 14.45$\pm$0.03 & ~42$\pm$~1.3     \\
0.199946$\pm$~1.2e-5$^a$ & ~83$\pm$11 & 13.17$\pm$0.06 & ~26$\pm$~4.6     &0.346824$\pm$~0.6e-5     & 221$\pm$16 & 13.59$\pm$0.02 & ~39$\pm$~1.9     \\
0.204831$\pm$~0.3e-5     & 208$\pm$~9 & 13.72$\pm$0.02 & ~24$\pm$~1.0     &0.348645$\pm$~0.9e-5     & ~40$\pm$10 & 12.78$\pm$0.06 & ~18$\pm$~3.0     \\
0.205326$\pm$~0.2e-5     & 322$\pm$~8 & 14.12$\pm$0.03 & ~25$\pm$~0.8   
\enddata
\end{tabular}
\vspace{-0.4cm} 
\footnotetext[1]{Due to blending, the uncertainties in
the line parameters could be larger than the formal uncertainties
estimated by VPFIT that are listed in this table.}
\footnotetext[2]{This line is part of a multicomponent fit, i.e., the
fit is not a single-component fit to an isolated line. Consequently,
there is a greater chance that the large Doppler parameter could be
due to a blend of several narrow components instead of thermal
broadening.}
\footnotetext[3]{For most high-$N$(H~I) lines, we have
fitted multiple Lyman series lines so that the fits are constrained by
adequately weak, unsaturated lines.  However, in this case, all
available lines are strongly saturated, and consequently the line
parameters are highly uncertain.}
\end{minipage}
\renewcommand{\thefootnote}{\arabic{footnote}}
\end{table*}
%}
%%%%%%%%%%%%%%%%%%%%%%%%%%%%%%%%%%%%%%%%%%%%%%%%%%%%%%%%%%%%%%%%%%%%%%%%%%%%%

%\clearpage

%\clearpage

\begin{figure}
\begin{center}
\resizebox{1.0\hsize}{!}{\includegraphics{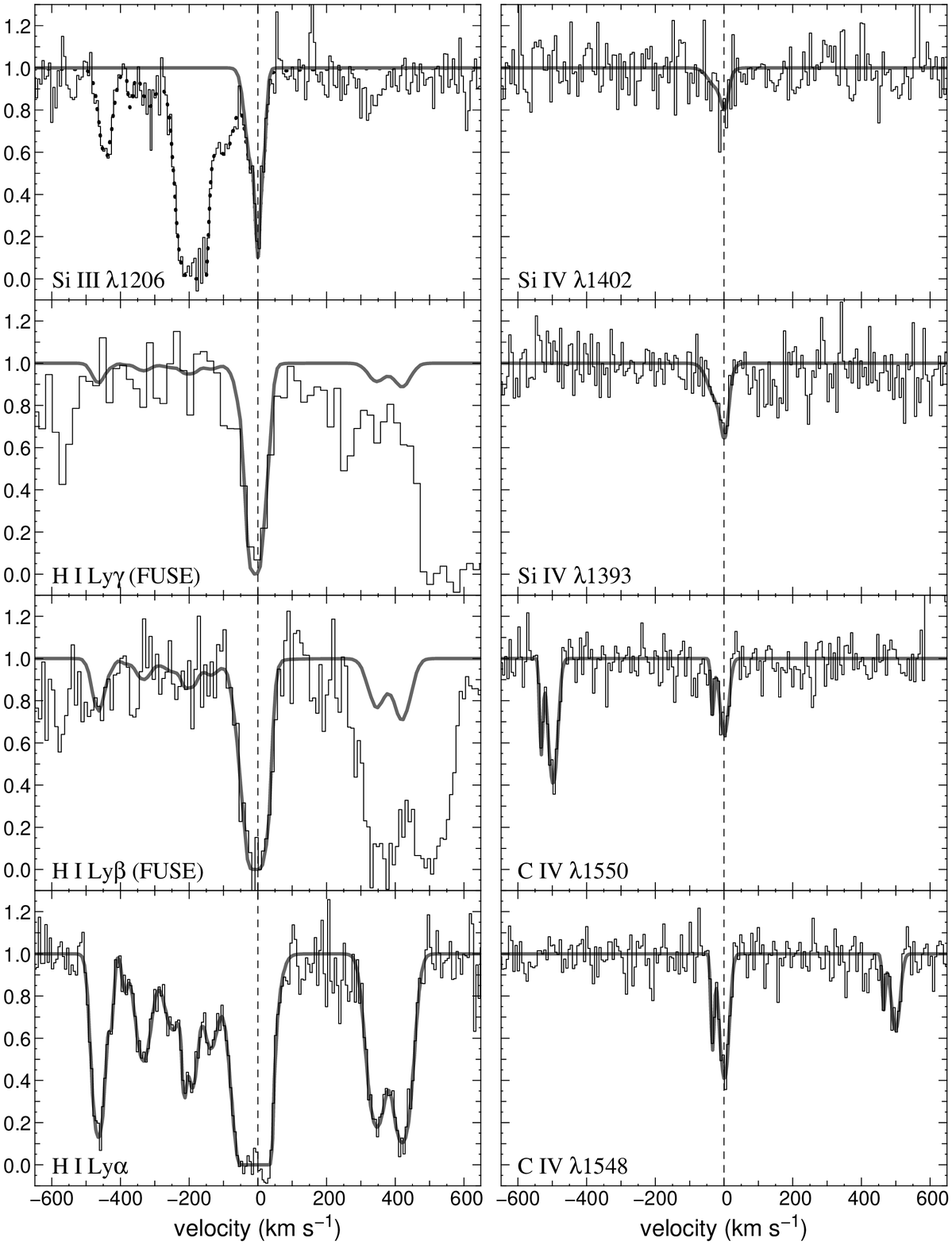}}
\caption{Transitions  observed at $z_{\rm  abs}$ = 0.06352
in the spectrum  of \hs0624 plotted in the absorber frame  ($v = 0$ km
s$^{-1}$ at $z_{\rm abs}$ =  0.06352).  The dashed line is centered on
``component A'', the strongest  \cion{H}{1} component in the Ly$\alpha$
cluster. The solid  line shows Voigt profiles fitted  to lines at this
redshift; the dotted  line indicates fits to unrelated  lines at other
redshifts.  These unrelated lines were  fitted in order to deblend the
features      from     the      \cion{Si}{3}      $\lambda     1206.50$
transition.}\label{fig:syst0063}
\end{center}
\end{figure}

Figure~\ref{fig:syst0063} compares the absorption profiles of the
Ly$\alpha$, Ly$\beta$, Ly$\gamma$, \ion{Si}{3}, \ion{Si}{4}, and
\ion{C}{4} lines at $z_{\rm abs}$ = 0.06352, and
Table~\ref{tab:cldnlist0064} lists the equivalent widths, column
densities, and $b-$values of the metals detected at this redshift as
well as upper limits on undetected species of interest.  Both
Voigt-profile fitting and direct integration of the ``apparent column
density'' profile (see Savage \& Sembach 1991; Jenkins 1996) were used
to estimate the metal column densities.  These independent methods are
generally in good agreement for the metal lines, which do not appear
to be strongly affected by unresolved saturation.  The \ion{Si}{3}
$\lambda 1206.50$ absorption at $z_{\rm abs} =$ 0.06352 is slightly
blended with a strong \ion{H}{1} line at $z_{\rm abs} =$ 0.05486 (see
Figure~\ref{fig:syst0063}). However, the metal lines at this redshift
have a distinctive two-component profile (see
Figures~\ref{fig:syst0063} and \ref{fig:civ064}), and the \ion{Si}{3}
profile shape is in good agreement with those of the \ion{C}{4} and
\ion{Si}{4} lines.  This indicates that (unrelated) blended \lya\ from
$z_{\rm abs}$ = 0.05486 contributes little optical depth to the
wavelength range where the \ion{Si}{3} absorption occurs. We fitted
the \ionl{Si}{3}{1206} and \ion{H}{1} \lya\ at $z_{\rm abs}=$0.05486 simultaneously, assuming
all of the \lya\ components are centered shortward of the \ion{Si}{3}
line.  The resulting joint fit is shown in Figure~\ref{fig:syst0063}.

%%%%%%%%%%%%%%%%%
%% System plot %%
%%%%%%%%%%%%%%%%%

%%%%%%%%%%%%%%%%%%%%%%%%
%% Curve of grow method %
%%%%%%%%%%%%%%%%%%%%%%%%
\begin{figure}
\begin{center}
\resizebox{1.0\hsize}{!}{\includegraphics{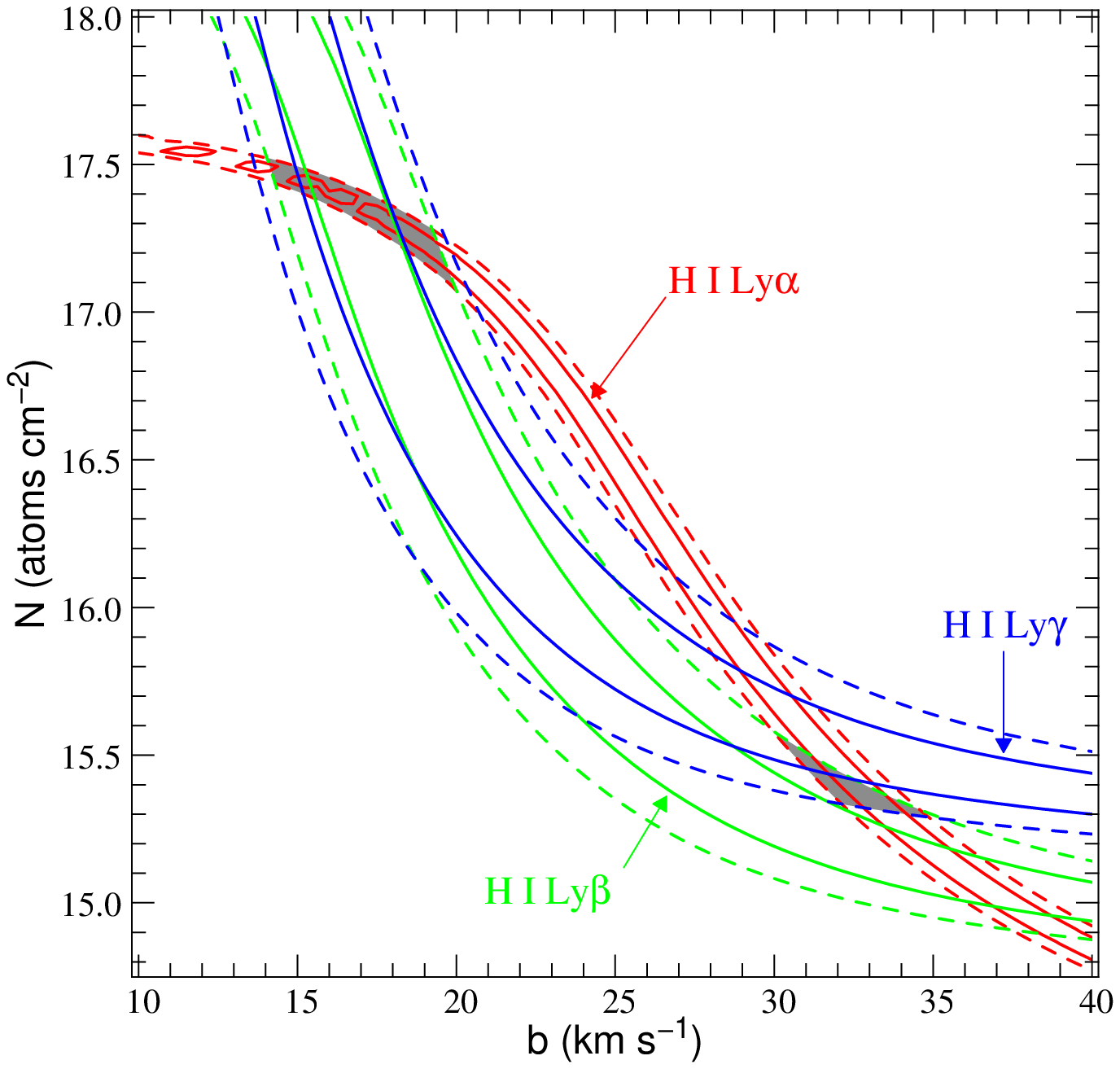}}
\caption{Contour maps of the equivalent width of \ion{H}{1} \lya\ (in
red), \lyb\ (in green) and \lyg\ (in blue) as a function of the column
density  $N$ and  the Doppler  parameter  $b$.  For  each transition,  the
contours   correspond   to   $W_{r}\pm1\sigma$  (solid   lines)   and
$W_{r}\pm2\sigma$ (dashed lines).  Only the values of $b$  and $N$ in the
shaded regions  can reproduce the  observed equivalent widths  for the
three transitions. }\label{fig:cog}
\end{center}
\end{figure}

\begin{figure}
\begin{center}
\resizebox{1.0\hsize}{!}{\rotatebox{-90}{\includegraphics{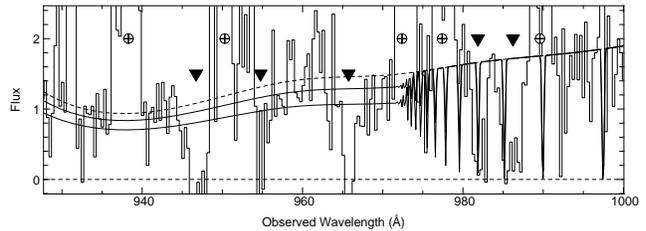}}}
\caption{Portion of the \hs0624 \FUSE\ spectrum. The data come from the
SiC2A channel and are rebinned over 10 pixels.  The crossed circles
indicate the position of the telluric lines and the black triangles
show the position of five strong H$_2$ absorption features.  The
dashed curve shows the adopted continuum placement, and the solid
curves represent two models of LLS absorption with
$N$(\ion{H}{1})=10$^{16.7}\;\cmmd$ (lower curve) and $10^{16.1}\;\cmmd$
(upper curve). }\label{fig:lls_rutr}
\end{center}
\end{figure}

The profile parameters derived for most of the \lya\ lines in the $z =
0.0635$   cluster  are  reasonably   well-constrained.  Some   of  the
components are  strongly blended  and are consequently  more uncertain
than the formal profile-fitting  error bars indicate; these are marked
in Table~\ref{tab:lyalist}. Component A  was also difficult to measure
for a different reason: the three usable Lyman series lines (Ly$\alpha$,
Ly$\beta$  and  Ly$\gamma$)  are  all saturated,  and  consequently  Voigt
profile fitting did not provide good constraints for the determination
of  the  component  A   \ion{H}{1}  column  densities.   As  shown  in
Figure~\ref{fig:cog}, using a single-component curve of growth and the
observed equivalent widths of  Ly$\alpha, \beta$, and $\gamma$, we
find  that  the  component   A  \ion{H}{1}  absorption  lines  can  be
reproduced by  two distinct sets  of values for the  \ion{H}{1} column
density  and the  Doppler  parameter.   One of  the  two sets  implies
N(\ion{H}{1})$\,\sim10^{17.4}\;\cmmd$,  which should produce  a strong
absorption edge  characteristic of  a Lyman Limit  System (LLS)  at an
observed wavelength of 970~\AA.   This wavelength region is covered by
the   \FUSE\  spectrum   in  the   SiC2a   channel  and   is  shown   in
Figure~\ref{fig:lls_rutr}.  The  S/N of the  SiC2a channel is  low but is
  adequate to constrain $N$(\ion{H}{1}).  The
optical depth  $\tau(\lambda)$ of the  Lyman limit absorption  and the
\ion{H}{1} column density are approximately related by

\begin{equation}
N({\rm H~I}) = \frac{\tau (\lambda)}{\sigma (\lambda)} = 1.6 \times
  10^{17}\left( \frac{912 \rm\AA}{\lambda}\right)^3\tau(\lambda)\;\;{\rm atoms \;\; cm}^{-2}\label{eq:nhills}
\end{equation}

\noindent where $\sigma (\lambda$) is the absorption cross section and
$\lambda$ is the rest-frame wavelength. The SiC2a spectrum does not
show any compelling evidence of a Lyman limit edge at the expected
wavelength, but the continuum placement is somewhat uncertain and
because of this, a small Lyman limit decrement could be present.
Based on the small depth of the flux decrement at $\lambda _{\rm obs}$
= 970 \AA , we derive a $3\sigma$ upper limit of $N$(\ion{H}{1}) $\leq
10^{16.1}$ cm$^{-2}$ (upper black solid curve in Figure
\ref{fig:lls_rutr}). We also show in Figure~\ref{fig:lls_rutr} the
Lyman limit absorption expected for $N$(\ion{H}{1}) = $10^{16.7}$
cm$^{-2}$ (lower black curve), which is too strong with our adopted
continuum placement. The absence of a strong Lyman limit edge rules
out the higher \ion{H}{1} column density of 10$^{17.5}$~cm$^{-2}$
predicted by the curve of growth shown in Figure~\ref{fig:cog}.  The
lower $N$(\ion{H}{1}) derived from the curve of growth
(10$^{15.4}$~cm$^{-2}$) is consistent with the lack of a Lyman limit
edge. To be conservative, we present below the metallicities derived
both from the upper limit [$N$(\ion{H}{1}) $\leq 10^{16.1}$] and from
the somewhat lower best value from the curve of growth shown in
Figure~\ref{fig:cog}.

%%%%%%%%%%%%%%%
% LLS figure %%
%%%%%%%%%%%%%%%
%
%\clearpage

\section{Ionisation and Metallicity}%
We next examine the physical conditions and metal enrichment of the
absorption systems implied by the column densities and the Doppler
parameters obtained from Voigt profile fitting. We concentrate on the
absorbers at $z_{\rm abs}$=0.06352 and 0.0757 because these systems show
metal absorption and can be associated with nearby galaxies/structures
as discussed in \S\ref{sec:environment}. To derive abundances from
the detected metals in these systems (\ion{Si}{3}, \ion{Si}{4}, and
\ion{C}{4}), we must apply ionisation corrections, which depend on the
ionisation mechanism and physical conditions of the gas.  We will show
that the gas is predominantly photoionised, and that the implied
metallicities are relatively high.

To investigate the absorber ionisation corrections and metallicities,
we employ CLOUDY photoionisation models (v96, Ferland et al. 1998) as
described in Tripp et al. (2003).  In these models, the absorbers are
approximated as constant density, plane-parallel gas slabs with a
thickness that reproduces the observed \ion{H}{1} column density.  The
gas in the cloud is photoionised by the UV background from quasars at
$z\simeq0.06$. We used the UV background spectrum shape computed by
Haardt \& Madau (1996) with the intensity normalized to
$J_\nu=1\times10^{-23}\;{\rm ergs\;s^{-1}\;cm^{-2}\;Hz^{-1}}$ at 1
Rydberg. This value is consistent with theoretical and observational
constraints (Shull et al. 1999; Weymann et al. 2001; Dav\'e \& Tripp
2001, and references therein).   With the models required to match the
observed $N$(\ion{H}{1}), the predicted metal column densities
depend mainly on the ionisation parameter $U$ ($\equiv$ ionising
photon density/total hydrogen number density), the overall
metallicity,\footnote{In this paper, we express metallicities using
the usual logarithmic notation, [X/Y] = log ($N$(X)/$N$(Y)) - log
(X/Y)$_{\odot}$.} and the assumed relative abundances of the
metals. We assume solar relative abundances, and we adopt recent
revisions reported by Allende Prieto et al.  (2001,2002) and Holweger
(2001) for oxygen, carbon, and silicon, respectively.

The high-ion column densities predicted by the photoionisation models
depend on the assumed UV background shape. In this paper, we primarily
use the UV background shape computed by Haardt \& Madau (1996), but we
note that other assessments of the UV background (e.g., Madau, Haardt,
\& Rees 1999; Shull et al. 1999) adopt a somewhat steeper EUV spectral
index for quasars, which changes the CLOUDY ionisation patterns for a
given metallicity and ionisation parameter.  We investigate how these
UV background uncertainties affect our results by modeling our systems
with both UV background shapes (see below).

% {\bf Actually, the predicted column densities are also affected by the choice of the UV background\
% shape too. As indicated previously we used the shape computed by Haardt \& Madau (1996) but this
% shape has been thereafter modified by changing the spectral index, $\alpha$, of the quasar spectra.
% Since the effects of the modification of $\alpha$ on  the UV background shape are different at different 
% wavelengths, this changes the predictions of the ionisation patterns for a given metallicity and 
% ionisation parameter.  We investigate thoses changes by modeling our systems with both UV 
% background shapes. ($\diamond$\it we need then to modify the part with the evaluation of the 
% metallicty but I cannot do that till the computation is done, so tomorrow.)}
 
 \subsection{$z_{\rm abs} = 0.06352$}\label{ss:z064}

{
\def~{\phn}
\def\fa{\footnote{Rest wavelengths and $f$-values are from Morton (1991).}}
\def\fb{\footnote{\parbox[t]{0.9\hsize}{Sum of the column densities obtained from Voigt profile fitting of the two evident components in the system.}}}
\def\fc{\footnote{Integrated apparent column density $N_a=\int N_a(v){\rm d}v$.}}
\begin{table}
\caption{Equivalents width and column densities of lines at $z\approx0.06352$\label{tab:cldnlist0064}}
\begin{minipage}{\hsize}
\begin{tabular}{@{}l@{ }c@{ }c@{  }ccc@{}}
   & $\lambda_0$ &  & $W_{obs}$ &   & \\
Species  & (\AA)   & $\log\,f\lambda_0$\fa & (m\AA)  & $\log\,N_{tot}$\fb & $\log\,N_{a}$\fc
\startdata
H~\textsc{i}   & 1215.670 & 2.704 & 603$\pm$~9 & 15.37$^{+0.10\mathstrut}_{-0.16}$ & \nodata \\
               & 1025.722 & 1.909 & 349$\pm$21 &  & \nodata\\
               & ~972.537 & 1.450 & 302$\pm$22 &  & \nodata\\
O \textsc{i}   & 1302.168 & 1.804 & ~~6$\pm$~9 &  & $<13.1$\\
O \textsc{vi}  & 1037.617 & 1.836 & ~28$\pm$17 &  & $<13.6$\\
C \textsc{ii}  & 1334.532 & 2.232 & 104$\pm$~9 &  & $<14.3$\\
C \textsc{iv}  & 1548.204 & 2.470 & 143$\pm$~9 & 13.67$\pm0.04$ & 13.61$\pm$0.03\\
               & 1550.781 & 2.169 & ~90$\pm$10 &  & 13.63$\pm$0.05\\
Si \textsc{ii} & 1260.422 & 3.104 & ~23$\pm$12 &  & $<12.8$\\
Si \textsc{iii}& 1206.500 & 3.304 & 151$\pm$10 & 13.02$^{+0.24}_{-0.21\mathstrut}$ & 13.08$\pm$0.03\\
Si \textsc{iv} & 1393.760 & 2.855 & ~76$\pm$10 & 13.05$^{+0.16}_{-0.12\mathstrut}$ & 12.95$\pm$0.06\\
               & 1402.773 & 2.554 & ~49$\pm$11 &  & 13.05$\pm$0.10\\
N \textsc{v}   & 1242.804 & 1.988 & ~23$\pm$10 &  & $<13.8$
\enddata
%\tablenotetext{a}{Rest wavelengths and $f$-values are from Morton (1991).}
%\tablenotetext{b}{Column densities obtained from Voigt profile fitting.}
%\tablenotetext{c}{Integrated apparent column density $N_a=\int N_a(v){\rm d}v$.}
\end{tabular}
\vspace{-6mm}
\end{minipage}
\end{table}
}

As  discussed  above,  the  \ion{H}{1} absorption  lines  detected  at
$z_{\rm  abs} =  0.0635$  are spread  over  a wide  velocity range  of
$\Delta v=1000$~\kms\ (see Figure~\ref{fig:hi00635}).   From the
velocity centroids of  the 13 fitted Voigt profiles and by using the biweight statistic as described in Beers et al. (1990),  we estimate that
the  line-of-sight velocity dispersion  of this  \ion{H}{1} absorption
complex  is  $\sigma_v=265$~\kms.   The  velocity  dispersion  of 
galaxies near  this redshift  is comparable to this value though  substantially more uncertain   (due   to  the   larger   uncertainties   in  the   galaxy redshifts). This  velocity dispersion is comparable  to those observed in elliptical-rich  galaxy groups  (e.g. Zabludoff \&  Mulchaey 1998),
which is interesting because elliptical-rich groups often show diffuse
X-ray emission (Mulchaey  \& Zabludoff 1998) indicative of  hot gas in
the   intragroup    medium.    However,    we   will   show    in   \S
\ref{sec:discussion} that the available information suggests that this
group is  predominantly composed of late-type spiral  and S0 galaxies.
Spiral-rich groups are much fainter  in X-rays but could still contain
hot intragroup gas if the gas is somewhat cooler ($10^{5} - 10^{6}$ K)
or has a much lower  density than that found in elliptical-rich groups
(Mulchaey 2000).  However, we argue that  most of the gas in the \lya\
complex at $z  = 0.0635$ is unlikely to  be hot, collisionally ionised
gas for several reasons:

First, the \ion{H}{1} lines in the \lya\ complex are generally too
narrow.  If the line broadening is dominated by thermal motions, then
the Doppler parameter is directly related to the gas temperature, $b =
\sqrt{2kT/m} = 0.129\sqrt{T/A}$, where $m$ is the mass, $A$ is the
atomic mass number, and the numerical coefficient is for $b$ in km
s$^{-1}$ and $T$ in K. Since other factors such as turbulence and
multiple components can contribute to the line broadening, $b-$values
provide only upper limits on the temperature.  Applying this equation
to the \lya\ line $b-$values from Table~\ref{tab:lyalist}, we find
that most of the \ion{H}{1} lines in the $z = 0.0635$ complex indicate
that $T \ll 10^{5}$~K, which is colder than expected for the diffuse
intragroup medium based on observed group velocity dispersions, even
in spiral-rich groups (e.g., Mulchaey et al. 1996). In a complex
cluster of \lya\ lines, it is easy to hide a broad \lya\ component
indicative of hot gas (see, e.g., Figure 6 in Tripp \& Savage 2000),
so the narrow \lya\ lines do not preclude the presence of hot gas, but
they do indicate that many cool clouds are present in the intragroup
medium.

\begin{figure}
\begin{center}
\resizebox{1.0\hsize}{!}{\includegraphics{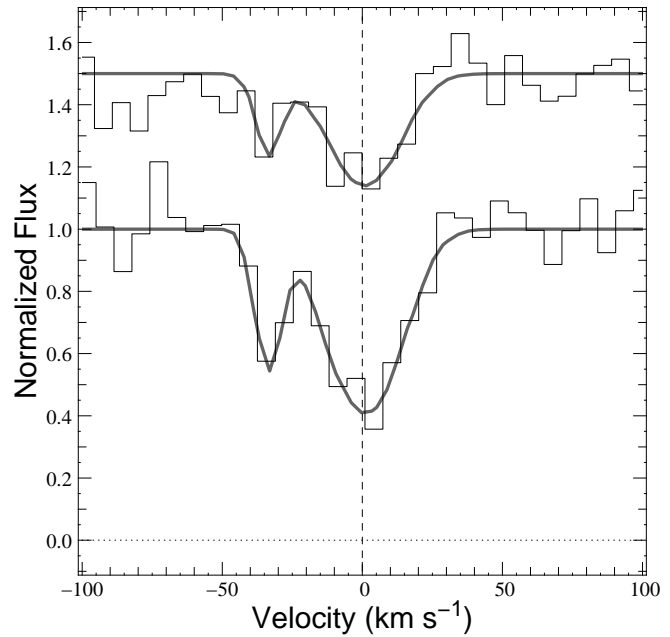}}
\caption{Expanded view of the continuum-normalized \ion{C}{4} $\lambda
  \lambda$1548.20, 1550.78 doublet at $z_{\rm abs} = 0.06352$, plotted
  vs. absorber velocity. In this figure, the data are binned 2 pixels
  into 1 (i.e., $\approx$ 7 km s$^{-1}$ pixels). The component at $-35$
  km s$^{-1}$ is formally found to have $b = 4\pm 2$ km s$^{-1}$, which
  is only marginally resolved at STIS E140M resolution. For clarity,
  the \ionl{C}{4}{1550.78} profile is shifted upward by
  0.5 flux units.\label{fig:civ064}}
\end{center}
\end{figure}

Second, the metal line profiles in component A favor cool,
photoionised gas. If component A metal lines were to arise in gas
in collisional ionisation equilibrium (CIE), the
$N$(\ion{C}{4})/$N$(\ion{Si}{4}) and $N$(\ion{Si}{4})/$N$(\ion{Si}{3})
 column density ratios (integrated across both components seen in these species, see Table \ref{tab:cldnlist0064}) would require a gas temperature $T \approx 10^{4.9}$~K (Sutherland \& Dopita 1993). However
the \ion{C}{4} component at $v=-35$~km~s$^{-1}$ is rather narrow. To show this, Figure \ref{fig:civ064} plots an expanded view of the \ion{C}{4} doublet. We see that the $v=-35$~km~s$^{-1}$ is marginally resolved at the STIS E140M resolution of $\sim7$~km~s$^{-1}$. Voigt profile fitting for this component formally yields $b \ =\;4\pm2$~\kms, which is significantly lower than the $b-$value implied by the CIE temperature, i.e., $b \approx 10.5$
km s$^{-1}$. The stronger component at $v = 0$ km s$^{-1}$ is broader
(see Figure~\ref{fig:civ064}), but the $N$(\ion{C}{4})/$N$(\ion{Si}{4}) and
$N$(\ion{Si}{4})/$N$(\ion{Si}{3}) ratios are similar in the components at $v =
-35$ and 0 km s$^{-1}$, and we expect the ionisation mechanism and
physical conditions to be similar in both components.

%%%%%%%%%%%%%%%%%%%%%%%%
% Figure CIV at z=0.064
%%%%%%%%%%%%%%%%%%%%%%%%
%%%%%%%%%%%%%%%%

%%%%%%%%%%%%%%%%
% Cloudy Model %
%%%%%%%%%%%%%%%%
\begin{figure}
\begin{center}
\resizebox{1.0\hsize}{!}{\includegraphics{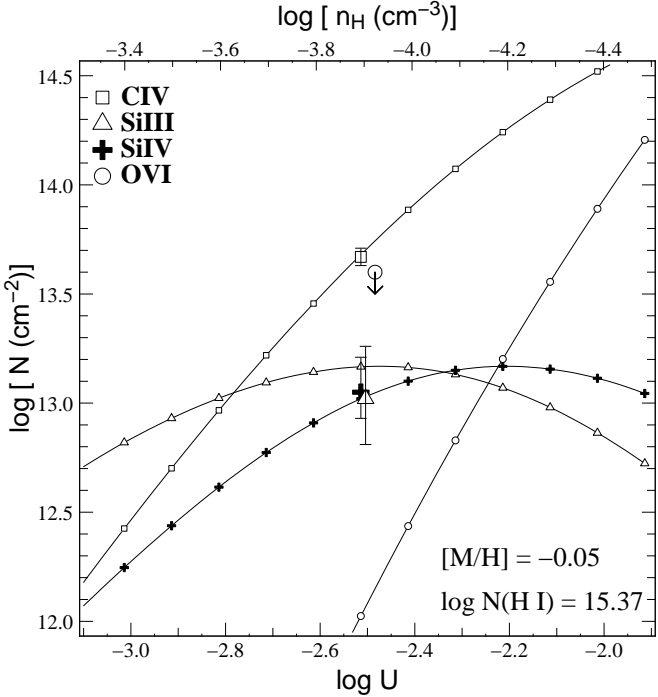}}
\caption{Predicted column densities, calculated with a CLOUDY
photoionisation model as described in the text, of \ion{O}{6} (small
circles), \ion{C}{4} (small squares), \ion{Si}{4} (small crosses), and
\ion{Si}{3} (small triangles) as a function of the ionisation
parameter $U$ (lower axis) and hydrogen number density $n_{\rm H}$
(upper axis) with $\log N$(\ion{H}{1})=15.37 and nearly solar
metallicity ([M/H] = $-0.05$). The larger markers with 1$\sigma$ error
bars are the observed \ion{C}{4}, \ion{Si}{4}, and \ion{Si}{3} column
densities measured from the STIS spectrum, plotted with the same
symbols at the best-fitting value of $U$. The 3$\sigma$ upper limit on
$N$(\ion{O}{6}) from the \FUSE\ data is indicated with a large open
circle.}\label{fig:cloudym}
\end{center}
\end{figure}
%%%%%%%%%%%%%%%%

Third, CLOUDY models photoionised by the UV background from QSOs are
fully consistent with the measured \ion{Si}{3}, \ion{Si}{4}, and
\ion{C}{4} column densities (and upper limits on undetected species)
at $z_{\rm abs}$ = 0.06352. Figure~\ref{fig:cloudym} shows the
relevant metal column densities predicted by CLOUDY models (small
symbols connected with solid lines) with log $N$(\ion{H}{1}) = 15.37
and [M/H] = $-0.05$ compared to the observed column densities (large
symbols). We can see that the metal column densities are in agreement
(within the 1$\sigma$ observational uncertainties) with this model at
log $U \approx -2.5$ (log $n_{\rm H} \approx -3.9$).

The narrow \ion{H}{1} and \ion{C}{4} components at this redshift could
still arise in shock-heated material if they originate in gas that is
not in ionisation equilibrium. Many papers have considered the
properties of gas that is initially shock-heated to some high
temperature and then cools more rapidly than it can recombine (e.g.,
Shapiro \& Moore 1976; Edgar \& Chevalier 1986).
However, if component A metal lines were to arise in gas in such a state, according to both computations from Shapiro \& Moore (1996) and Schmutzler \& Tscharnuter (1992),  the $N$(\ion{Si}{3})/$N$(\ion{Si}{4})  column density ratio would require a gas temperature similar ($T \approx 10^{4.9}$~K) to the one found for the collisional ionisation equilibrium hypothesis. Moreover, assuming solar abundances, the predicted \ion{O}{6} column density at this temperature  is always higher than our observed upper limit ($\sim$5 times higher in the Schmutzler model). Finally, the $N$(\ion{C}{4})/$N$(\ion{Si}{4}) column density ratio implies an even higher temperature than 10$^{4.9}$~K (2.5 times higherr in the Shapiro model). Because of these points, this non-equilibrium cooling gas scenario seems unlikely
 to apply to the $z_{\rm abs} = 0.06352$ absorber toward \hs0624.

%  Such a scenario could lead to highly ionised gas that yet is cool enough to satisfy the line
%width constraints presented above. However, if component A metal lines were
% arise in gas in such state, the $N$(\ion{C}{4})/$N$(\ion{Si}{4}) and $N$(\ion{Si}{3})/$N$(\ion{Si}{4})
% column density ratios would require a gas temperature similar ($T \approx 10^{4.9}$~K) than
%  the one found for the collisional ionisation equilibrium hypothesis.
%   This non-equilibrium cooling gas scenario is therefore unlikely
%  to apply to the $z_{\rm abs} = 0.06352$ absorber toward \hs0624

%  However, models of such highly ionised,
%non-equilibrium cooling gas predict strong \ion{O}{6} absorption lines
%(e.g., Edgar \& Chevalier 1986; Heckman et al. 2002). This
%non-equilibrium cooling gas scenario is therefore unlikely to apply to
%the $z_{\rm abs} = 0.06352$ absorber toward \hs0624 because no
%\ion{O}{6} is evident at the expected level.

The CLOUDY modeling has some other interesting implications in
addition to the basic conclusion that the gas is photoionised. For
example, { the photoionisation model indicates that the absorber has a
relatively high metallicity of $Z\simeq0.9Z_\odot$ even though we have found no luminous
galaxies within $\rho\leq135$~\hmkpc.} A similarly high
metallicity ([O/H]$\simeq-0.2$) was recently reported by Jenkins et al. (2005) for a LLS
in the spectrum of PHL1811, but that system is much closer in
projection to a luminous galaxy. If we adopt the more conservative
upper limit on $N$(\ion{H}{1}) from the absence of a Lyman limit edge
($N$(\ion{H}{1})$=10^{16.1}$~\cmmd, see Figure~\ref{fig:lls_rutr}) instead of the curve-of-growth
\ion{H}{1} column, we obtain [M/H] $\geq -0.75$. This lower limit is
still substantially higher than metallicities typically observed in
analogous absorbers at higher redshifts (e.g., Schaye et al. 2003) and
is comparable to abundances seen in high-velocity clouds near the
Milky Way (e.g., Sembach et al. 2001,2004; Collins, Shull, \& Giroux
2003; Tripp et al. 2003; Ganguly et al. 2005; Fox et al. 2005).

To derive confidence limits on parameters extracted from our CLOUDY
models, for combinations of metallicity $Z$ and ionisation parameter
$U$ we calculated the $\chi^{2}$ statistic,
\begin{equation}
\chi^2(Z,U) = \sum_i\Big(\frac{N_{i,obs} -
N_{i,model}(Z,U)}{\sigma(N_{i,obs})}\Big)^2,
\end{equation}
where $N$ indicates column density and the sum is over the three ions
\ion{Si}{3}, \ion{Si}{4}, and \ion{C}{4}. With the minimum $\chi^{2}$
obtained at [M/H] = $-0.05$ and log $U = -2.5$, we evaluated
confidence limits by finding parameters that increased $\chi ^{2}$ by
the amount appropriate for a given confidence level (see Lampton,
Margon, \& Bowyer 1976; Press et al. 1992).  In this way, we find
[M/H] = $-0.05\pm 0.4$ at the 2$\sigma$ confidence level. Of course,
these confidence limits do not fully reflect potential sources of
systematic error such as uncertainties in the shape of the ionising
flux field or accuracy of the atomic data incorporated into
CLOUDY. 
When we used the steeper UV background shape (e.g. Madau, Haardt, \& Rees 1999; Shull et al. 1999), the observations are still consistent with the CLOUDY model for a lower metallicity of [M/H] = $-0.24$ and a larger ionisation parameter log $U = -2.4$.

%  The variation of the metallicity induced by the uncertainties in the shape of the ionising flux is then lower than the confidence levels we estimated with the $\chi^{2}$ method.}
%Assessment of such errors is beyond the scope of this paper,
%but we expect that potential systematics are unlikely to increase the
%errors by more than a few tenths of a dex. 

\begin{figure}
\begin{center}
\resizebox{1.0\hsize}{!}{\includegraphics{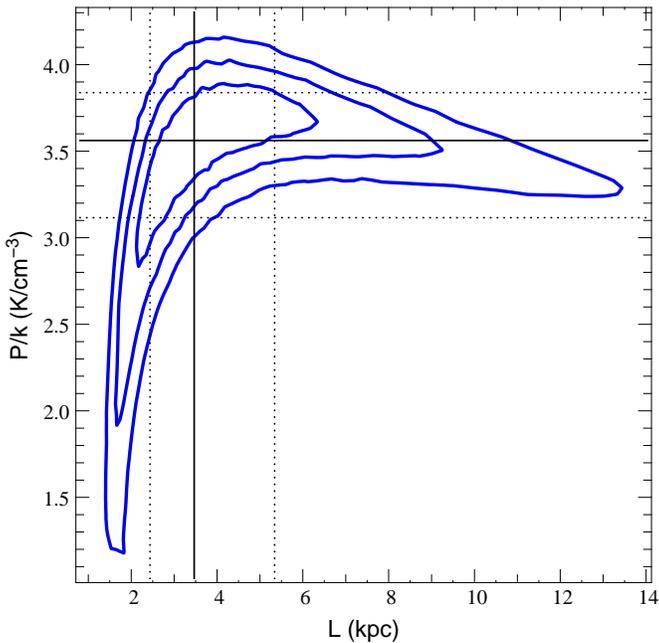}}
\caption{The thermal pressure $P/k$ of the absorber  vs.   its  size $L$ 
parameter contours  at 60,  90 and 99 per cent  confidence levels  in the
joint-fit with the CLOUDY  photoionisation modelling. The solid and
the dotted black lines indicate for each parameter the best estimation
and          the          68.7 per cent        confidence          interval
respectively.\label{fig:absorbersize}}
\end{center}
\end{figure}

We  can also  place constraints  on  physical quantities  such as  the
absorber size  (the length of  the path through the  absorbing region)
and the thermal gas pressure (but see the caveats discussed in \S 5 of
Tripp  et al.  2005).  Figure~\ref{fig:absorbersize} shows  confidence
interval contours for the absorber size $L$ and thermal pressure $P/k$
implied by  the photoionisation model.   The best fit implies  that $L
\approx$ 3.5 kpc. If spherical,  the baryonic mass of this cloud would
be   $\approx   10^{5}   M_{\odot}$.   However,  we   can   see   from
Figure~\ref{fig:absorbersize} that the model  allows a large range for
$L$ at the 2$\sigma$ level.   The low thermal pressure implied is also
notable.  When the steeper UV background shape is used for the CLOUDY model, the predicted pressure is lower by a factor 1.5 and the absorber size increases to $\simeq$8~kpc.%
The   range   of   pressures  within   the   contours   in
Figure~\ref{fig:absorbersize}  is several  orders  of magnitude  lower
than  the gas  pressure measured  in the  disk of  the Milky  Way (see
Jenkins \&  Tripp 2001) and is  even lower than  pressures measured in
HVCs in  the Milky Way halo  (e.g. Wakker, Oosterloo,  \& Putman 2002;
Fox et al.  2005). However, Sembach et al. (1995, 1999) found similar pressure for CIV HVCs surrounding the Milky-Way with somewhat lower metallicity. 
Moreover, pressures this low  are predicted in some
theoretical models of galactic  halos (Wolfire et al. 1995). Finally,
the derived pressure depends on the intensity  used to normalize
 the ionising flux field (see  Tripp et al. 2005) and both  the particle density and
the pressure could  be higher if the radiation  field is brighter than
we assumed.

%However, pressures this low  are predicted in some
%theoretical models of galactic  halos (Wolfire et al. 1995). Moreover,
%the derived pressure depends on the intensity assumed for the ionising
%flux field (see  Tripp et al. 2005) and both  the particle density and
%the pressure could  be higher if the radiation  field is brighter than
%we assumed.

\subsection{$z_{\rm abs} = 0.07573$}\label{ss:z076}

\begin{figure}
\begin{center}
\resizebox{1.0\hsize}{!}{\includegraphics{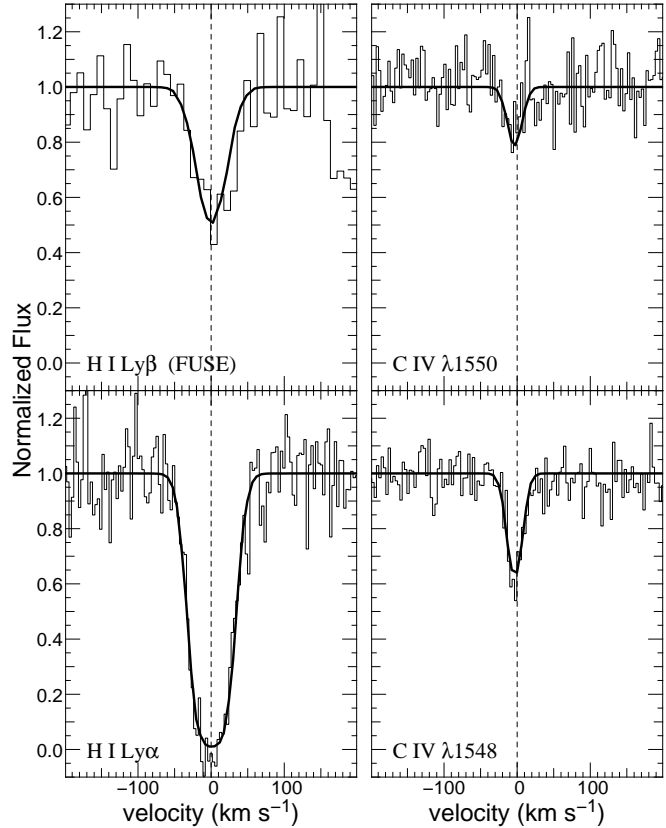}}
\caption{Transitions observed at $z_{\rm abs}$ = 0.07573 in the
spectrum of \hs0624 including \ion{H}{1} Ly$\alpha$, Ly$\beta$, and
the \ion{C}{4} doublet. The continuum-normalized profiles are plotted
vs. velocity in the absorber frame ($v$ = 0 km s$^{-1}$ at $z_{\rm
abs}$ = 0.07573). }\label{fig:syst0076}
\end{center}
\end{figure}

As discussed in \S \ref{sec:environment}, the absorption lines
detected at $z_{\rm abs}$ = 0.07573 occur in a large-scale structure
connected to \abell{559} and \abell{564}. Only \ion{H}{1} Ly$\alpha$,
Ly$\beta$ and the \ion{C}{4} $\lambda \lambda$1548.20,1550.78 doublet are
clearly detected at this redshift; the profiles of these absorption
lines are shown in Figure~\ref{fig:syst0076}.  The weakness of
Ly$\beta$ and the absence of Ly$\gamma$ suggest that the \ion{H}{1}
lines are not badly affected by unresolved saturation, and profile
fitting measurements are robust. Likewise, comparison of the
\ion{C}{4} apparent column density profiles shows no evidence of
unresolved saturation. We have fitted these lines with only one
component.  The results of the fit are listed in Table
\ref{tab:cldnlist076}. The width of the \ion{H}{1} line ($b=24.6$~\kms)  indicates a
temperature for the gas lower than $10^{4.5}$~K, which again favors a
photoionisation process. No \ion{O}{6} is evident at this redshift,
but several strong unrelated lines of various elements are found close to the expected wavelength of
the \ion{O}{6} doublet, and these lines might mask weak \ion{O}{6}
absorption.

Despite the fact that \ion{C}{4} is the only metal detected in this
system, we can nevertheless place an interesting lower limit on the
absorber metallicity.  The carbon abundance can be expressed as [C/H]=
log[\Nion{C}{4}/\Nion{H}{1}] + log[\fcion{H}{1}/\fcion{C}{4}] - log (C/H)$_\odot$, where \fcion{H}{1} and \fcion{C}{4} are the ion fractions of
\ion{H}{1} and \ion{C}{4}, respectively\footnote{We can derive a lower limit because \fcion{H}{1}/\fcion{C}{4} has a minimum value or, put another way, \Nion{C}{4} has a maximum value for any \Nion{H}{1} and [C/H] combination. The maximum \Nion{C}{4} is not evident in Figure~\ref{fig:cloudym} because it occurs at a higher value of $U$ than the range shown.}. With the \ion{H}{1} column
from the Ly$\alpha$+Ly$\beta$ fit (Table \ref{tab:cldnlist076}), and
again assuming that the gas is photoionised by the UV background from
QSOs (Haardt \& Madau 1996), we find that
log[\fcion{H}{1}/\fcion{C}{4}]$\geq-3.2$, and therefore
$[{\rm C}/{\rm H}]>-0.6$ in the \zabs{0.07573} absorber.
Once again, this metallicity lower limit is relatively high despite
the fact that no luminous galaxies have been found near the sight line
(the closest galaxy is NW3 at $\rho \ = 293 h_{70}^{-1}$ kpc, see
Table~\ref{tab:spec_red}). 
 
%%%%%%%%%%%%%%%%%%%%%%
% Stack plot z=0.076 %
%%%%%%%%%%%%%%%%%%%%%%
%%%%%%%%%%%%%%%%%%%%%%

%%%%%%%%%%%%%%%%%%%%
% Eqw system 0.076 %
%%%%%%%%%%%%%%%%%%%%
{
\def~{\phn}
\begin{table}
\begin{minipage}{\hsize}
\caption{Equivalents width and column densities of detected lines at $z = 0.07573$ \label{tab:cldnlist076}}
\begin{tabular}{@{}l@{}c@{\hspace{3ex}}c@{\hspace{3ex}}c@{\hspace{3ex}}c@{\hspace{3ex}}c@{}}
\colhead{} & \colhead{$\lambda_0$} & \colhead{$W_{obs}$} & \colhead{} & \colhead{$b$} & \colhead{}\\
\colhead{Species}  & \colhead{(\AA)}  &  \colhead{(m\AA)} & \colhead{log $N_{tot}$} & \colhead{\kms} & \colhead{log $N_{a}$}

\startdata
H~\textsc{i}   & 1215.670 & 309$\pm$10 & 14.18$\pm$0.04 & 24.6$\pm$1.0 & 14.06$\pm$0.03\\
               & 1025.722 & 108$\pm$21 &                &              & 14.28$\pm$0.08\\
O~\textsc{vi}  & 1037.617 & ~~0$\pm$19 &                &              & $<13.9$\\
C~\textsc{iv}  & 1548.204 & ~51$\pm$~7 & 13.18$\pm$0.04 & 13.1$\pm$1.5 & 13.24$\pm$0.04\\
               & 1550.781 & ~27$\pm$~8 &                &              & 12.09$\pm$0.22\\
Si~\textsc{iii}& 1206.500 & ~~0$\pm$10 & & & $<12.1$\\
Si~\textsc{iv} & 1393.760 & ~17$\pm$14 & & & $<12.8$\\
N~\textsc{v}   & 1238.821 & ~25$\pm$~9 & & & $<13.4$

\enddata
\end{tabular}
\end{minipage}
\end{table}
}
%%%%%%%%%%%%%%%%%%%
%\clearpage

\subsection{$z_{\rm abs} = 0.113$}\label{ss:z113}

In \S  \ref{sec:environment} we  noted that the  clusters \abell{554},
\labell{562},   and   \labell{565}   indicate   that   a   large-scale
filament/supercluster  is foreground to  \hs0624. Our  galaxy
redshift survey  has also revealed  four galaxies close to  \hs0624 at
the    redshift   of    the   Abell    554/562/565    structure   (see
Figure~\ref{fig:field}  and Table~\ref{tab:spec_red}),  which suggests
that   the   large-scale   filament   extends   across   the   \hs0624
field. However,  we only find a  couple of weak  Ly$\alpha$ lines near
the  redshift of  this structure at redshifts
substantially  offset from  those  of the  galaxies  and Abell
clusters.  It is possible  that the  Ly$\alpha$ lines  are weak/absent
because the  gas in  the filament  is so hot  that the  \ion{H}{1} ion
fraction makes the Ly$\alpha$ line undetectable, but we also note that
the  nearest galaxies are  farther from  the sight  line in  this case
($\rho \geq  1.2 h_{70}^{-1}$ Mpc)  than in the  structures at $z  = $
\zva\ and \zvb\ discussed above, so it is also possible that the sight
line does not penetrate the part of the dark matter filament where the
potential  is  deep  enough   to  accumulate  gas  and  galaxies  (see
discussion in Bowen et al. 2002).

\section{Comparison to Other Ly$\alpha$ Absorbers}

How do the properties of the Ly$\alpha$ absorbers at \zva , \zvb , and
\zvc\ compare to the other Ly$\alpha$ lines in the \hs0624 spectrum
(and in other sight lines)?  We have found that the systems at \zva\
and \zvb\ arise in photoionised cool gas; is this true of the majority
of the Ly$\alpha$ lines in the spectrum?  In particular, do we find
Ly$\alpha$ lines that arise in hot gas? Richter et al. (2004, 2005)
and Sembach et al. (2004) have recently identified a population of
broad Ly$\alpha$ lines (BLAs) with $b >$ 40 km s$^{-1}$ in the spectra
of several low$-z$ QSOs (PG0953+415, PG1116+215, PG1259+593, and
H1821+643).  Williger et al. (2005) similarly find a substantial
number of BLAs in the spectrum of PKS0405-123. Bowen et al. (2002) have also identified some BLA candidates using somewhat lower resolution data. If these lines are
mainly broadened by thermal motions, then they trace the warm-hot IGM,
and moreover, they in this case would contain a substantial portion of
the baryons in the Universe at the present epoch (see Richter et
al. 2004, 2005; Sembach et al. 2004).  Based on simulations, Richter
et al. (2005) and Williger et al. (2005) find that some of the BLAs
are not predominantly thermally broadened but instead are due to line
blends that are difficult to recognize at the S/N afforded by typical
STIS echelle spectra.  However, Richter et al. (2005) conclude that
approximately 50 per cent of the BLAs are mainly thermally broadened, and
some high S/N examples in the above papers are remarkably smooth and
broad and appear to entirely consistent with a single broad Gaussian
(see, e.g., Figures 4 and 5 in Richter et al. 2005).

\begin{figure}
\begin{center}
\resizebox{1.0\hsize}{!}{\includegraphics{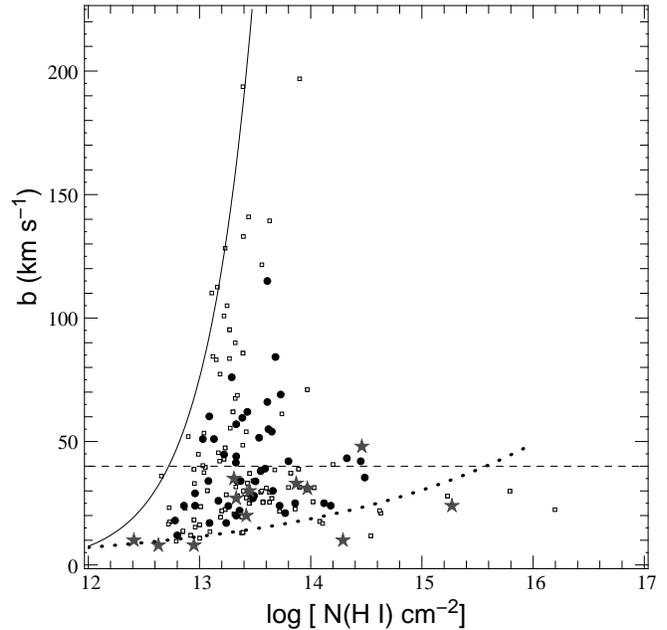}}
\caption{Doppler parameters $b$ as a function of the \ion{H}{1} column
density $N$ for the Ly$\alpha$ lines detected in the spectrum of
\hs0624 (solid circles, see Table~\ref{tab:lyalist}). The Ly$\alpha$
lines detected in the complex at $z$ = \zva\ are shown with stars. For
comparison, we also plot the measurements from Sembach et al. (2004)
and Richter et al. (2004) from the PG1116+215 and PG1259+593 sight
lines (open squares). The solid line is the relation between $b$ and
$N$ for a Ly$\alpha$ absorption line with central optical depth = 0.1;
this represents a detection threshold for these data.  The dotted line
shows the minimum $b-$value as a function of $N$(\ion{H}{1}) predicted
by equation 5 of Dav\'{e} \& Tripp (2001), which is based on the
hydrodynamic cosmological simulations of Dav\'{e} et al. (1999). Lines
above the horizontal dashed line have $T > 10^{5}$ K if the lines are
predominantly thermally broadened.}\label{fig:larg}
\end{center}
\end{figure}

In this paper, in addition to using a different sight line, we have
employed methods that are independent from (e.g., using different
software) the techniques used in the papers above for continuum
normalization, line detection, and profile fitting. Consequently, we
have an opportunity to independently check the BLA findings reported
in these papers.  For the \hs0624 sight line, we find that the mean
$b-$value for all Ly$\alpha$ lines is $<b>$ = 37 km s$^{-1}$, and the
median $b_{m} = 33$ km s$^{-1}$. However, as noted in
Table~\ref{tab:lyalist}, some of the Ly$\alpha$ lines are
significantly blended, and the line parameters are accordingly
uncertain.  If we exclude these uncertain blended cases, we find $<b>
= 34$ km s$^{-1}$ and $b_{m} = 30$ km s$^{-1}$. These ensemble
$b-$values are in reasonable agreement with the previous
high-resolution Ly$\alpha$ studies.

In Figure~\ref{fig:larg} we compare our measurements of the \ion{H}{1}
$b-$values and column densities from Table~\ref{tab:lyalist} to the
measurements reported by Richter et al. (2004) and Sembach et
al. (2004). The solid line indicates $b$ vs. $N$(\ion{H}{1}) for a
Gaussian line with central optical depth = 0.1.  This is effectively a
detection threshold; lines that have ($b,N$) combinations to the left
of this line are not likely to be detected.  The dotted line in
Figure~\ref{fig:larg} shows the minimum $b-$value as a function of
$N$(\ion{H}{1}) predicted by Dav\'{e} \& Tripp (2001, see their
equation 5) from the hydrodynamic cosmological simulations of Dav\'{e}
et al. (1999). This predicted lower envelope appears to be in
reasonable agreement with the observed lower envelope for the three
sight lines shown in the figure. 

From Figure~\ref{fig:larg}, we see that the ($b,N$) distribution that
we have obtained from the \hs0624 sight line appears to be generally
similar to those obtained by Sembach et al. (2004) and Richter et
al. (2004). Sembach et al. and Richter et al. did find more extremely
broad Ly$\alpha$ lines ($b >$ 80 km s$^{-1}$) than we have been able
to positively identify in the \hs0624 spectrum.  This may be partly
due to signal-to-noise differences -- the data employed by Richter et
al. and Sembach et al. have higher S/N -- because such broad and
shallow lines are difficult to detect in the \hs0624 data.  For $40 <
b < 80$ km s$^{-1}$, the different sight lines appear to be in broad
agreement. Excluding lines within 5000 km s$^{-1}$ of the QSO
redshift,\footnote{Lines within 5000 km s$^{-1}$ of the QSO redshift
can arise in intrinsic gas ejected by the QSO, and these lines can be
rather broad (see, e.g., Yuan et al. 2002), even if not part of a
full-blown broad absorption line outflow.} the \hs0624 spectrum can be
used to search for BLAs between $z_{\rm min} = 0.004$ and $z_{\rm max}
= 0.347$. Accounting for regions in which broad lines could have been
masked by IGM or ISM lines, we obtain a blocking-corrected total
redshift path $\Delta z = 0.329$.  With 21 Ly$\alpha$ lines in the
sample with $b >$ 40 km s$^{-1}$, we thus obtain $dN/dz$(BLA) =
64$\pm$16. This is somewhat larger than the values reported by Sembach
et al. (2004) and Richter et al. (2004,2005).  However, Richter et
al. have excluded BLAs that are located in complex blends on the
grounds that these cases are more likely to be affected by non-thermal
broadening.  If we follow the same procedure, we must reject 10 BLAs
(see Table~\ref{tab:lyalist}); the remaining 11 BLAs would then imply
$dN/dz$(BLA) = 33$\pm$10.  Using equations 1, 5, and 6 from Sembach et
al. (2004), but adjusted for the somewhat different cosmological
parameters assumed in this paper, we find that our full sample implies
that the BLA baryonic content is $\Omega_{b}$(BLA) = 0.017
$h_{70}^{-1}$ (in the usual notation, i.e., $\Omega =
\rho/\rho_{c}$). This high value probably substantially overestimates
the BLA baryonic content, largely because of false BLAs that arise
from blends. If we exclude BLAs that are located in complex blends,
this drops to $\Omega_{b}$(BLA) = 0.0036 $h_{70}^{-1}$, which is
similar to values obtained by Richter et al. and Sembach et al.  The
uncertainties in $\Omega_{b}$(BLA) due to, e.g., lines that are broad
due to blends or other non-thermal broadening mechanisms, are large
and currently difficult to assess (see discussion in Richter et
al. 2005).  However, these initial calculations suggest that BLAs may
harbor an important quantity of baryons. With future UV spectrographs,
it would be valuable to obtain high-resolution spectra with
substantially better S/N in order to accurately assess the baryonic
content of BLAs as part of the general census of ordinary matter in
the nearby Universe.

\section{Discussion}\label{sec:discussion}

We have acquired detailed information about the abundances, physical
conditions, and galaxy proximity of absorption systems in the
direction of \hs0624.  What are the implications of these measurements
for broader questions of galaxy evolution and cosmology?  The
processes that add gas to galaxies (e.g. accretion) and remove gas from galaxies
(e.g., winds, dynamical stripping) can have profound effects on galaxy
evolution, and the ``feedback'' of matter and energy from galaxies
into the IGM is now believed to play an important
role in shaping structures that subsequently grow out of the IGM (Voit G. M., 2005).  The quantity and implications of the $10^{5} - 10^{7}$ K WHIM gas is a topic of
particular interest currently.  The galaxies and absorption systems in
the direction of \hs0624, particularly the galaxy group and Ly$\alpha$
complex at $z$ = \zva , have some interesting, and perhaps surprising,
implications regarding these questions, which we now discuss.

\begin{figure}
\begin{center}
\resizebox{1.0\hsize}{!}{\includegraphics{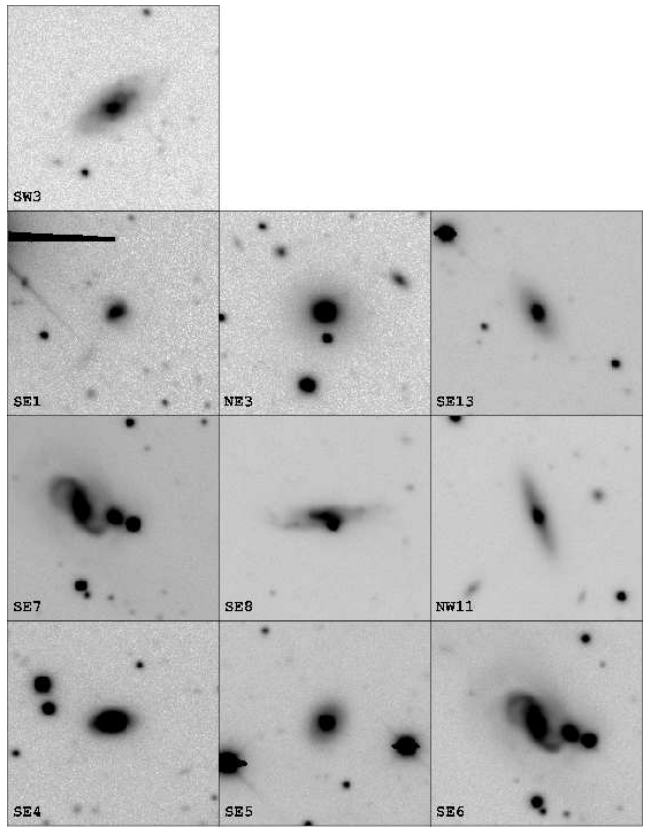}}
\caption{Images of galaxies found in the $z$ = \zva\ group in the
field of \hs0624, recorded in the $R$ band with the KPNO 4m MOSA camera.
Each box in the montage
spans $50''\times 50''$, and the galaxy name from
Table~\ref{tab:spec_red} is listed in the lower left corner. The
galaxy is in the center of each panel. SE6 and SE7 are in close
proximity; SE6 is the more extended spiral galaxy.}\label{fig:galpics}
\end{center}
\end{figure} 

ROSAT observations of diffuse X-ray emission have established that
galaxy groups that are dominated by early-type galaxies often contain
diffuse, hot intragroup gas (Mulchaey 2000, and references
therein). Based on the observed relation between intragroup gas
temperature and velocity dispersion $\sigma$ in X-ray bright groups
($T \propto \sigma ^{2}$) and the fact that spiral-rich groups have
lower velocity dispersions than elliptical-rich groups, Mulchaey et
al. (1996) have hypothesized that spiral-rich groups might have
somewhat cooler intragroup media that could give rise to QSO
absorption lines at WHIM temperatures (e.g., \ion{O}{6}).  

However, the galaxy group at $z$ = \zva\ appears to have properties
that are not consistent with the elliptical-rich groups detected with
ROSAT nor with the idea that spiral-rich groups contain warm-hot
intragroup gas. It is unclear if the galaxy group at $z$ = \zva\ is a
spiral-rich group.  Figure~\ref{fig:galpics} shows $R$ images from the
MOSA data of the 10 galaxies that we have found in this group. Most of
the galaxies in the group show evidence of disks and bulges (both in
the direct images and in radial brightness profiles). We find from
visual inspection that 4-5 of the 10 galaxies have indications of
spiral structure (SW3, SE1, SE8, SE6, and possibly SE4). However, the
remaining galaxies could be early-type S0 galaxies, and therefore the
early-type fraction might be comparable to groups that show diffuse
X-ray emission (see, e.g, Figure 7 in Zabludoff \& Mulchaey 1998). The
NE3, SE13, and SE5 galaxies, which morphologically appear to be
early-type galaxies, have colours and magnitudes consistent with the
``red sequence'' colour-magnitude relation observed in clusters (e.g.,
Bower, Lucey, \& Ellis 1992; McIntosh et al. 2005); these galaxies are
likely S0s (the other 7 galaxies have blue colours characteristic of
late types). The velocity dispersion of the group at $z$ = \zva ,
albeit uncertain, is more comparable to those of elliptical-dominated
groups than spiral-rich groups (Mulchaey et al. 1996; Zabludoff \&
Mulchaey 1998).

Regardless of whether the group is elliptical- or spiral-rich, it is
surprising that we find a large number of cool, photoionised clouds in
the intragroup medium (\S \ref{ss:z064}).  In the hot intragroup
medium of an elliptical-rich group, \ion{H}{1} lines should be
extremely broad and weak, but instead we find strong, narrow lines
(see Figure~\ref{fig:hi00635}).  Even in the cooler gas predicted to
be found in late-type dominated groups, the \ion{H}{1} lines should be
broader. We could entertain models of cooling intragroup gas, but in
such models \ion{O}{6} is expected to be stronger.  Likewise, if the
intragroup gas is a multiphase medium with cooler clouds (which cause
the \ion{H}{1} absorption lines) embedded in a hotter phase, then we
might expect to detect \ion{O}{6} from the interface between the
phases (Fox et al. 2005), unless conduction is somehow suppressed.

The lack of evidence of hot gas leads us to question whether this
group is a bound, virialized system.  An alternative possibility is
that our sight line passes along the long axis of a large-scale
filamentary structure in the cosmic web. In this case, the projection
of the galaxies and Ly$\alpha$ clouds along the sight line could give
a false impression of a group in which hot gas would be
expected. However, in cosmological simulations of large-scale
filaments, WHIM gas is expected to be widespread at the present epoch,
even in modest-overdensity regions (see, e.g., Figure 4 in Cen \&
Ostriker 1999b), so it is interesting that we find a substantial
number of cool clouds at $z \approx$ \zva , somewhat contrary to
theoretical expectations. As noted above, our data do not preclude the
presence of WHIM gas at $z \approx$ \zva , but we find no clear
evidence of it. Other sight lines show similar clusters of Ly$\alpha$
lines, e.g., the Ly$\alpha$ complex at $z_{abs} \approx$ 0.057 toward
PKS2155-304 (Shull et al. 1998; Shull, Tumlinson, \& Giroux 2003) or
the Ly$\alpha$ lines at $z_{abs} \approx$ 0.121 toward H1821+643
(Tripp et al. 2001).  However, unlike the \hs0624 Ly$\alpha$ complex,
the PKS2155-304 and H1821+643 examples both show evidence of warm-hot
intragroup gas. To test whether the observations and simulations are
in accord, it would be useful to assess the frequency and physical
properties of these Ly$\alpha$ complexes in cosmological simulations
for comparison with the observations.

It is also interesting that the two systems for which we have obtained
abundance constraints (at $z$ = \zva\ and \zvb ) both indicate
relatively high metallicities, but both of these systems are at least
100 kpc away (in projection) from the nearest known galaxy. This
naturally raises a question: how did gas that is so far from a galaxy
attain such a high metallicity?  The gas could have been driven out of
a galaxy by a galactic wind; some wind models predict that the
outflowing material will have a high metallicity (Mac Low \& Ferrara
1999), The difficulty with this interpretation is that winds from
nearby galaxies are usually observed to contain substantial amounts of
hot gas (e.g., Strickland et al. 2004), which seems to be inconsistent
with the absorption line properties as we have discussed.  

A more likely explanation is that the high-metallicity gas we have
detected in absorption has been tidally stripped out of one of the
nearby galaxies. There are indications that tidal stripping could be a
more gentle process for removing gas from galaxies, and a tidally
stripped origin can therefore more easily accommodate the observed
low-ionisation state of the gas. For example, in the direction of
NGC3783, the Galactic high-velocity cloud (HVC) at $v_{\odot} =$ 247
\kms\ is now recognized to be tidally stripped debris from the
SMC. While this tidally stripped material shows a wide array of
low-ionisation absorption lines, it has little or no associated
high-ion absorption (Lu et al. 1994; Sembach et al. 2001a).  Moreover,
the tidally-stripped HVC contains molecular hydrogen, which Sembach et
al. argue formed in the SMC and survived the rigors of tidal stripping
(as opposed to forming in situ in the stream). Both the absence of
high ions and the survival of H$_{2}$ suggest that the stripping
process did not substantially ionise and heat this HVC. Several
galaxies are close enough to the \hs0624 sight line to be plausible
sources of the gas in a tidal stripping scenario. One of the nearby
galaxies, SE1, has a distorted spiral morphology.  This galaxy is a
plausible source of tidally stripped matter.

\section{Summary}

We have presented a study of absorption-line systems in the direction
of \hs0624 using a combination of high-resolution UV spectra obtained
with {\it HST}/STIS and \FUSE\ plus ground-based imaging and
spectroscopy of galaxies within $\sim$30' of the sight line.  In
addition to presented the basic measurements and ancillary
information, we have reported the following findings:

1. There are several Abell galaxy clusters in the foreground of
   \hs0624, including two clusters at $z \approx$ 0.077 (\abell{559} and
   \labell{564}) and three at $z \approx$ 0.110 (\abell{554}, \labell{562}, and
   \labell{565}). These clusters trace large-scale dark matter structures,
   i.e., superclusters or filaments of the ``cosmic web''.  Our galaxy
   redshift survey has revealed galaxies at these supercluster
   redshifts in the immediate vicinity of the \hs0624 sight line, and
   therefore our QSO spectra provide an opportunity to study the gas
   in large-scale intergalactic filaments. The most prominent group of
   galaxies found in our galaxy redshift survey is at $z \approx$
   \zva\ and is not associated with an Abell cluster with a
   spectroscopic redshift from the literature. However, \abell{557}, for
   which no spectroscopic redshift has been reported, is at least
   partly due to the galaxy group at $z \approx$ \zva .

2. The two strongest Ly$\alpha$ absorption systems at $z < 0.2$ arise
   in galaxy groups at $z = $ \zva\ and \zvb .  The Ly$\alpha$
   absorption at \zva\ is particularly dramatic: at this redshift, we
   find 13 Ly$\alpha$ lines spread over a velocity range of 1000 km
   s$^{-1}$ with a line-of-sight velocity dispersion of 265 km
   s$^{-1}$.  The second-strongest system at \zvb\ is associated with
   the Abell 559/564 large-scale structure, and this indicates that a
   filament containing gas and galaxies feeds into the Abell 559/564
   supercluster.

3. Analysis of the Ly$\alpha$ absorption-line complex at $z =$ \zva\
   provides strong evidence that the gas is photoionised and
   relatively cool; we find no compelling evidence of warm-hot gas in
   this large-scale filament.  We detect \ion{Si}{3}, \ion{Si}{4}, and
   \ion{C}{4} in the strongest component in this Ly$\alpha$ complex,
   and photoionisation models indicate that the gas metallicity is
   high, [M/H] = $-0.05 \pm 0.3$.  This is somewhat surprising because
   we do not find any luminous galaxies close to the sight line; the
   closest galaxy is at a projected distance $\rho = 135 h_{70}^{-1}$
   kpc.  The Ly$\alpha$ system at \zvb\ is only detected in the
   \ion{C}{4} doublet, but nevertheless we find a similar result: the
   lower limit on the metallicity is relatively high ([C/H] $> -0.6$)
   while the nearest galaxy is at $\rho = 293 h_{70}^{-1}$ kpc.

4. We have compared the distribution of Ly$\alpha$ Doppler parameters
   and \ion{H}{1} column densities to high-resolution measurements
   obtained from other sight lines, and we find good agreement.  We
   find that the number of broad Ly$\alpha$ absorbers with $b >$ 40 km
   s$^{-1}$ per unit redshift is in agreement with results recently
   reported by Richter et al. (2004,2005) and Sembach et al. (2004).
   The baryonic content of these broad \ion{H}{1} lines is still
   highly uncertain and requires confirmation with higher S/N data,
   but it is probable that some of these broad line arise in warm-hot
   gas and contain an important portion of the baryons in the nearby
   universe.  We also find that the lower bound on $b$
   vs. $N$(\ion{H}{1}) is in agreement with predictions from
   cosmological simulations.

5. The absence of warm-hot gas in the galaxy group/Ly$\alpha$ complex
   at $z =$ \zva\ is difficult to reconcile with X-ray observations of
   bound galaxy groups.  It seems more likely that this in this case
   we are viewing a large-scale cosmic web filament along its long
   axis.  The filament contains a mix of early- and late-type galaxies
   and many cool, photoionised clouds.

6. The high-metallicity, cool cloud at $z_{\rm abs}$ = 0.06352 is
   probably tidally stripped material.  This origin can explain the
   high metallicity and the lack of hot gas. One of the nearby
   galaxies has a disturbed morphology consistent with this
   hypothesis.

Similar clusters of Ly$\alpha$ lines have been observed in other sight
lines, and additional examples are likely to be found as we continue
to analyse STIS data.  We look forward to comparisons of these
observations to predictions from cosmological simulations and other
theoretical work in order to better understand the processes that
affect the evolution of galaxies and the intergalactic medium.

\section*{Acknowledgments}

We thank Dan McIntosh and Neal Katz for useful discussions. The STIS
observations of HS0624+6907 were obtained for {\it HST} program 9184
with financial support through NASA grant HST GO-9184.08-A from the
Space Telescope Science Institute. This research was also supported in
part by NASA through Long-Term Space Astrophysics grant NNG
04GG73G. The \FUSE\ data were obtained by the PI team of the
NASA-CNES-CSA \FUSE\ project, which is operated by Johns Hopkins
University with financial support through NASA contract NAS
5-32985. This research has made use of the NASA/IPAC Extragalactic
Database (NED), which is operated by the Jet Propulsion Laboratory,
California Institute of Technology, under contract with NASA.

% @ARTICLE{1991ApJ...379..245S,
%    author = {{Savage}, B.~D. and {Sembach}, K.~R.},
%     title = "{The analysis of apparent optical depth profiles for interstellar absorption lines}",
%   journal = {\apj},
%      year = 1991,
%     month = sep,
%    volume = 379,
%     pages = {245-259},
%    adsurl = {http://adsabs.harvard.edu/cgi-bin/nph-bib_query?bibcode=1991ApJ...379..245S&db_key=AST},
%   adsnote = {Provided by the NASA Astrophysics Data System}
% }

\end{document}